\documentclass[preprint]{elsarticle}

\usepackage{microtype}
\usepackage{amsmath}
\usepackage{color}
\usepackage{pifont}
\usepackage{enumitem}
\usepackage{amssymb}
\usepackage{tikz}

\usepackage{subcaption}
\usepackage{polylib}

\newtheorem{theorem}{Theorem}
\newtheorem{lemma}[theorem]{Lemma}

\newenvironment{proof}{{\noindent \em Proof:~}}{\hfill{\hfill$\Box$}}

\graphicspath{{./graphics/}}
\polypath{./graphics/}

\newcommand*{\figref}[1]{\figurename~\ref{#1}}
\newcommand*{\figrefs}[1]{Figures~\ref{#1}}
\newcommand*{\myeqref}[1]{Equation~\eqref{#1}}
\newcommand{\abs}[1]{\left|#1\right|}
\newcommand{\N}{\mathbb{N}}
\newcommand{\Z}{\mathbb{Z}}
\newcommand{\hex}{\mathcal{H}}
\newcommand{\squ}{\mathcal{S}}
\newcommand{\set}[1]{\left\lbrace#1\right\rbrace}
\newcommand{\bra}[1]{\left[#1\right]}
\newcommand{\ceil}[1]{\left\lceil#1\right\rceil}
\newcommand{\floor}[1]{\left\lfloor#1\right\rfloor}

\newcommand\polylog{{\rm polylog}}

\newcommand{\perim}[1]{\mathcal{P}(#1)}
\newcommand{\border}[1]{\mathcal{B}(#1)}
\newcommand{\minp}{\epsilon}
\newcommand{\minn}{\minp^{-1}}
\newcommand{\psize}{e_P(Q)}
\newcommand{\bsize}{e_B(Q)}
\newcommand{\poly}{Q}
\newcommand{\lattice}{\mathcal{L}}
\newcommand{\hexagon}{\drawpolyhex[scale=0.6]{single_hex.txt}}

\newcommand{\myqed}{}  
\newcommand{\comment}[1]{\relax}
\newcommand{\caplabel}[1]{#1}

\newif\ifusestandalone
\usestandalonetrue 
\ifusestandalone
	\usepackage{standalone}
\fi

\begin{document}

\title{Minimal-Perimeter Lattice Animals and the Constant-Isomer Conjecture}

\author[tech1]{Gill~Barequet\corref{cor}}
\ead{barequet@cs.technion.ac.il}
\author[tech1]{Gil~Ben-Shachar}
\ead{gilbe@cs.technion.ac.il}
\cortext[cor]{Corresponding author}
\address[tech1]{
   Center for Graphics and Geometric Computing,
   Dept.\ of Computer Science, \\
   The Technion---Israel Inst.\ of Technology,
   Haifa 3200003, Israel.
}


\begin{abstract}
   We consider minimal-perimeter lattice animals, providing a set of conditions which are sufficient
   for a lattice to have the property that inflating all minimal-perimeter animals of a
   certain size yields (without repetitions) all minimal-perimeter animals of a new, larger
   size.  We demonstrate this result on the two-dimensional square and hexagonal lattices.
   In addition, we characterize the sizes of minimal-perimeter animals on these lattices that are not
   created by inflating members of another set of minimal-perimeter animals.
\end{abstract}


\maketitle


\begin{center}
\fbox{\includegraphics[scale=0.79]{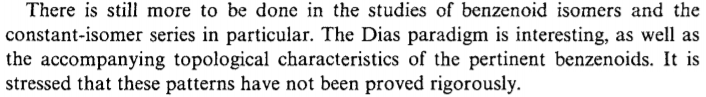}}
\vspace{-0.5mm} \\
\begin{minipage}{4.7in}
   \footnotesize
   Cyvin S.J., Cyvin B.N., Brunvoll J. (1993) Enumeration of benzenoid chemical isomers with
   a study of constant-isomer series. In: \emph{Computer Chemistry}, part of \emph{Topics in
   Current Chemistry} book series, vol.\ 166.  Springer, Berlin, Heidelberg (p.~117).
\end{minipage}
\end{center}


\section{Introduction}

An \emph{animal} on a $d$-dimensional lattice is a connected set of lattice cells, where 
connectivity is through ($d{-}1$)-dimensional faces of the cells.  Specifically, on the planar
square lattice, connectivity of cells is through edges.  Two animals are considered identical if
one can be obtained from the other by \emph{translation} only, without rotations or flipping.
(Such animals are called ``fixed'' animals, as opposed to ``free'' animals.)

\begin{figure}
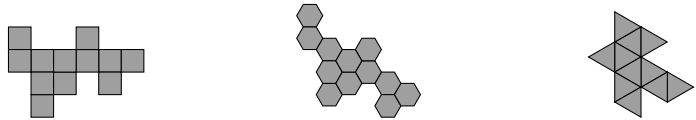

    \centering
    \begin{subfigure}[t]{0.3\textwidth}
    \centering
    \drawpoly[scale = 0.75]{exmpSqr.txt}
    \end{subfigure}
    \begin{subfigure}[t]{0.3\textwidth}
    \centering
    \drawpolyhex[scale = 0.75]{exmpHex.txt}
    \end{subfigure}
    \begin{subfigure}[t]{0.3\textwidth}
    \centering
    \drawpolyiamond[scale = 0.70]{exmpTri.txt}
    \end{subfigure}
    \caption{An example of a polyomino, a polyhex, and a polyiamond.}
    \label{fig:examples}
\end{figure}

Lattice animals attracted interest in the literature as combinatorial objects~\cite{eden1961two} and
as a computational model in statistical physics and chemistry~\cite{temperley1956combinatorial}.
(In these areas, one usually considers \emph{site} animals, that is, clusters of lattice
vertices, hence, the graphs considered there are the \emph{dual} of our graphs.)
In this paper, we consider lattices in two dimensions, specifically, the hexagonal, triangular,
and square lattices, where animals are called polyhexes, polyiamonds, and
polyominoes, respectively.
We show the application of our results to the square and hexagonal lattices,
and explain how to extend the latter to the triangular lattice.
An example of such animals is shown in figure~\ref{fig:examples}.
Let $A^\lattice(n)$ denote the number of lattice animals of size~$n$, that is, animals
composed of $n$ cells, on a lattice~$\lattice$.
A major research problem in the study of lattices is understanding the nature
of~$A^\lattice(n)$, either by finding a formula for it as a function of~$n$, or by evaluating
it for specific values of~$n$.
These problems are to this date still open for any nontrivial lattice.
Redelmeier~\cite{redelmeier1981counting} introduced the first algorithm for
counting all polyominoes of a given size, with no polyomino being generated more than once.
Later, Mertens~\cite{Mertens1990} showed that Redelmeier's algorithm can be
utilized for any lattice.
The first algorithm for counting lattice animals without generating all of them was
introduced by Jensen~\cite{jensen2000statistics}.  Using his method, the number of animals on
the 2-dimensional square, hexagonal, and triangular lattices were computed up to size~56,
46, and~75, respectively. 

An important measure of lattice animals is the size of their \emph{perimeter} (sometimes called
``site perimeter'').  The perimeter of a lattice animal is defined as the set of empty cells
adjacent to the animal cells.  This definition is motivated by percolation models in
statistical physics.  In such discrete models, the plane or space is made of small cells
(squares or cubes, respectively), and quanta of material or energy ``jump'' from a cell to
a neighboring cell with some probability.  Thus, the perimeter of a cluster determines
where units of material or energy can move to, and guide the statistical model of the flow.

\begin{figure}
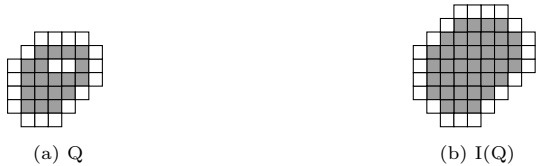

    \centering
    \begin{subfigure}[t]{0.45\textwidth}
    \centering
    \drawpoly[scale = 0.45]{exmpQ.txt}
    \caption{Q}
    \end{subfigure}
    \begin{subfigure}[t]{0.45\textwidth}
    \centering
    \drawpoly[scale = 0.45]{exmpIQ.txt}
    \caption{I(Q)}
    \end{subfigure}
    \caption{A polyomino~$\poly$ and its inflated
            polyomino~$I(\poly)$.  Polyomino cells are colored gray, while
            perimeter cells are colored white.}
    \label{fig:exmp}
\end{figure}
Asinowski et al.~\cite{asinowski2017enumerating,asinowski2018polycubes} provided
formulae for polyominoes and polycubes with perimeter size close to the maximum possible.
On the other extreme reside animals with the \emph{minimum} possible perimeter size for their area.
The study of polyominoes of a minimal perimeter dates back to Wang and Wang~\cite{wang1977discrete},
who identified an infinite sequence of cells on the square lattice, the first~$n$ of
which (for any~$n$)
form a minimal-perimeter polyomino.  Later, Altshuler et
al.~\cite{altshuler2006}, and independently Sieben~\cite{sieben2008polyominoes}, studied
the closely-related problem of the \emph{maximum} area of a polyomino with~$p$ perimeter cells,
and provided a closed formula for the minimum possible perimeter of $n$-cell polyominoes.

Minimal-perimeter animals were also studied on other lattices.
For animals on the triangular lattice (polyiamonds), the main result is due to
F\"{u}lep and Sieben~\cite{fulep2010polyiamonds}, who characterized all the polyiamonds
with maximum area for their perimeter, and provided a formula for the minimum perimeter of a
polyiamond of size~$n$.
Similar results were given by Vainsencher and Bruckstein~\cite{VainsencherB08}
for the hexagonal lattice.
In this paper, we study an interesting property of minimal-perimeter animals, which relates to the
notion of the \emph{inflation} operation.  Simply put, inflating an animal is adding to it
all its perimeter cells (see Figure~\ref{fig:exmp}).
We provide a set of conditions (for a given lattice), which if it holds , then inflating all
minimal-perimeter animals of some size yields all minimal-perimeter animals of some larger size in a
bijective manner.

While this paper discusses some combinatorial properties of minimal-perimeter polyominoes,
another algorithmic question emerges from these properties, namely,
``how many minimal-perimeter polyominoes are there of a given size?''
This question is addressed in detail in a companion paper~\cite{barequet2020algorithms}.

The paper is organized as follows.
In Section~\ref{sec:main}, we provide some definitions and prove our main theorem.
In sections~\ref{sec:polyominoes} and~\ref{sec:polyhexes}, we show the application of
Section~\ref{sec:main} to polyominoes and polyhexes, respectivally.
Then, in Section~\ref{sec:polyiamonds} we explain how the same result also applies to the
regular triangular lattice.
We end in Section~\ref{sec:conclusion} with some concluding remarks.

\subsection{Polyhexes as Molecules}

In addition to research of minimal-perimeter animals in the literature on combinatorics,
there has been much more intensive research of minimal-perimeter polyhexes in the literature on organic
chemistry, in the context of the structure of families of molecules.
For example, significant amount of work dealt with molecules called \emph{benzenoid hydrocarbons}.
It is a known natural fact that molecules made of carbon atoms are structured as shapes on the
hexagonal lattice.  Benzenoid hydrocarbons are made of
carbon and hydrogen atoms only.  In such a molecule, the carbon atoms are arranged as a
polyhex, and the hydrogen atoms are arranged around the carbons atoms.

\begin{figure}
   \centering
   \begin{tabular}{ccc}
      \raisebox{0.39\height}{\includegraphics[scale=0.13]{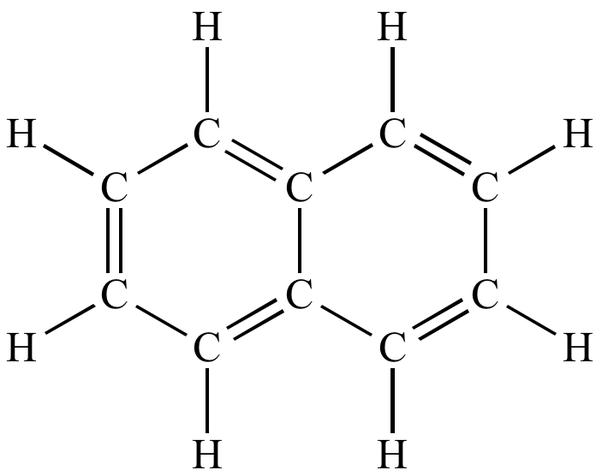}} & ~~~ &
         \includegraphics[scale=0.13]{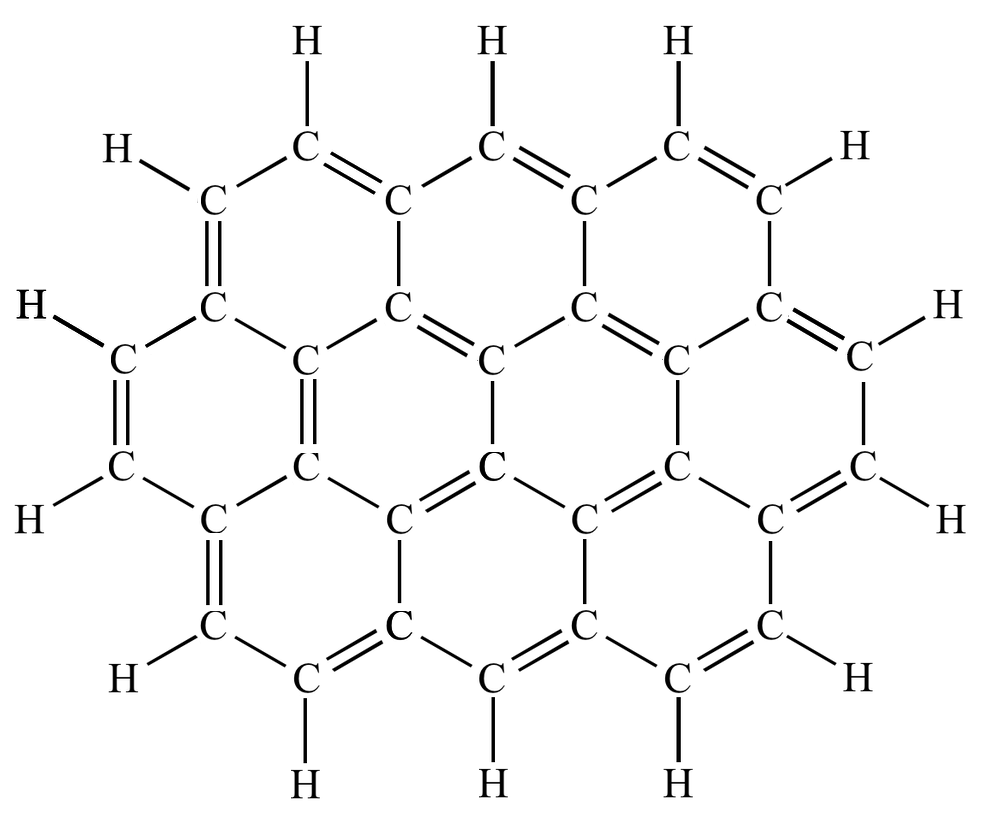} \\
      (a) Naphthalene ($C_{10} H_8$) & & (b) Circumnaphtaline ($C_{32} H_{14})$
   \end{tabular}
   \caption{Naphthalene and its circumscribed version.}
   \label{fig:naphthalene}
\end{figure}
Figure~\ref{fig:naphthalene}(a) shows a schematic drawing of the molecule of Naphthalene
(with formula~$C_{10}H_8$), the simplest benzenoid hydrocarbon, which is made of ten carbon
atoms and eight hydrogen atoms, while Figure~\ref{fig:naphthalene}(b) shows Circumnaphthalene
(molecular formula~$C_{32}H_{14}$).
There exist different configurations of atoms for the same molecular formula, which are
called \emph{isomers} of the same formula.
In the field of organic chemistry, a major goal is to enumerate all the different
isomers of a given formula.
Note that the carbon and hydrogen atoms are modeled by lattice \emph{vertices} and not by
cells of the lattice, but as we explain below, the numbers of hydrogen atoms identifies with the
number of perimeter cells of the polyhexes under discussion.
Indeed, the hydrogen atoms lie on lattice vertices that do not belong to the polyhex formed by the
carbon atoms (which also lie on lattice vertices), but are connected to them by lattice edges.
In minimal-perimeter polyhexes, each perimeter cell contains exactly two such hydrogen vertices,
and every hydrogen vertex is shared by exactly two perimeter cells.
(This has nothing to do with the fact that a single cell of the polyhex might be neighboring
several---five, in the case of Naphthalene---``empty'' cells.)
Therefore, the number of hydrogen atoms in a molecule of a benzenoid hydrocarbon is identical to the
size of the perimeter of the imaginary polyhex.\footnote{
   In order to model atoms as lattice cells, one might switch to the dual of the hexagonal lattice,
   that is, to the regular triangular lattice, but this will not serve our purpose.
}

In a series of papers (culminated in Reference~\cite{dias1987handbook}), Dias provided the
basic theory for the enumeration of benzenoid hydrocarbons.
A comprehensive review of the subject was given by Brubvoll and Cyvin~\cite{brunvoll1990we}.
Several other works~\cite{harary1976extremal,cyvin1991series,dias2010} also dealt with the
properties and enumeration of such isomers.
The analogue of what we call the ``inflation'' operation is called \emph{circumscribing} in
the literature on chemistry.
A circumscribed version of a benzenoid hydrocarbon molecule~$M$ is created by adding to~$M$
an outer layer of hexagonal ``carbon cells,'' that is, not only the hydrogen atoms (of~$M$)
adjacent to the carbon atoms now turn into carbon atoms, but also new carbon atoms are added
at all other ``free'' vertices of these cells so as to ``close'' them.
In addition, hydrogen atoms are put at all free lattice vertices that are connected by edges
to the new carbon atoms.
This process is visualized well in Figure~\ref{fig:naphthalene}.
In the literature on chemistry, it is well known that circumscribing all isomers of a given
molecular formula yields,
in a bijective manner,
all isomers that correspond to another molecular formula.
(The sequences of molecular formulae that have the same number of isomers created by
circumscribing are known as \emph{constant-isomer series}.)
Although this fact is well known, to the best of our knowledge, no rigorous proof of it
was ever given.

As mentioned above, we show that inflation induces a bijection between sets of
minimal-perimeter animals on the square, hexagonal, and in a sense, also on the triangular lattice.
By this, we prove the long-observed (but never proven)
phenomenon of ``constant-isomer series,'' that is, that circumscribing isomers of benzenoid
hydrocarbon molecules (in our terminology, inflating minimum-perimeter polyhexes)
yields all the isomers of a larger molecule.


\section{Minimal-Perimeter Animals}
\label{sec:main}

Throughout this section, we consider animals on some specific lattice~$\lattice$.
Our main result consists of a set of conditions on minimal-perimeter animals on~$\lattice$,
which is sufficient for satisfying a bijection between sets of minimal-perimeter animals on~$\lattice$.

\subsection{Preliminaries}
\label{subsec:preliminaries}

\begin{figure}
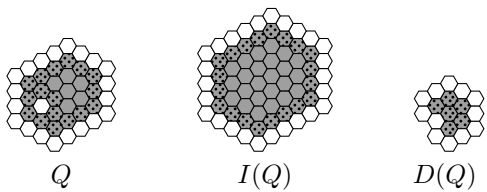

   \centering
   \begin{tabular}{ccccc}
      \drawpolyhex[scale=0.5]{exmp_hex.txt} & &
      \drawpolyhex[scale=0.5]{exmp_hex_I.txt} & &
      \drawpolyhex[scale=0.5]{exmp_hex_D.txt}\\
      $Q$ & & $I(Q)$ & & $D(Q)$
   \end{tabular}
   \caption{A polyhex~$\poly$, its inflated polyhex~$I(\poly)$, and its
            deflated polyhex~$D(\poly)$.  The gray cells belong to~$\poly$,
            the white cells are its perimeter, and
            its border cells are marked with a pattern of dots.}
    \label{fig:exmp_poly}
\end{figure}
Let~$\poly$ be an animal on~$\lattice$.
Recall that the \emph{perimeter} of~$\poly$, denoted by~$\perim{\poly}$, is the set of all
empty lattice cells that are neighbors of at least one cell of~$\poly$.
Similarly, the \emph{border} of~$\poly$, denoted by~$\border{\poly}$, is the set of cells
of~$\poly$ that are neighbors of at least one empty cell.
The \emph{inflated} version of~$\poly$ is defined as~$I(\poly) := \poly \cup \perim{\poly}$.
Similarly, the \emph{deflated} version of~$\poly$ is defined
as~$D(\poly) := \poly \backslash \border{\poly}$.
These operations are demonstrated in Figure~\ref{fig:exmp_poly}.

Denote by~$\minp(n)$ the minimum possible size of the perimeter of an $n$-cell
animal on~$\lattice$,
and by~$M_n$ the set of all minimal-perimeter $n$-cell animals on~$\lattice$.

\subsection{A Bijection}

\begin{theorem}
   \label{thm:main}
   Consider the following set of conditions.
   \begin{enumerate}[label=(\arabic*)]
      \item The function~$\minp(n)$ is weakly-monotone increasing.
      \item There exists some constant~$c^* = c^*(\lattice)$, for which, for any minimal-perimeter
            animal~$\poly$, we have that~$\abs{\perim{\poly}} = \abs{\border{\poly}} + c^*$
            and~$\abs{\perim{I(\poly)}} \leq \abs{\perim{\poly}}+c^*$.
      \item If~$\poly$ is a minimal-perimeter animal of size $n+\minp(n)$, then~$D(\poly)$ is a
            valid (connected) animal.
   \end{enumerate}
   If all the above conditions hold for~$\lattice$,
   then~$\abs{M_n} = \abs{M_{n+\minp(n)}}$.
   If these conditions are not satisfied for only a finite amount of sizes of animals, then
   the claim holds for all sizes greater than some lattice-dependent nominal size~$n_0$.
   \myqed
\end{theorem}

\begin{proof}  
   We begin with proving that inflation preserves perimeter minimality.

   \begin{lemma}
      \label{lemma:minimal-inflating}
      If~$\poly$ is a minimal-perimeter animal, then~$I(\poly)$ is a
      minimal-perimeter animal as well.
   \end{lemma}

   \begin{proof}
      Let~$\poly$ be a minimal-perimeter animal. Assume to the contrary
      that~$I(\poly)$ is not a minimal-perimeter animal, thus, there exists
      an animal~$\poly'$ such that $\abs{\poly'} = \abs{I(\poly)}$,
      and~$\abs{\perim{\poly'}} < \abs{\perim{I(\poly)}}$.
      By the second premise of Theorem~\ref{thm:main}, we know that
      $\abs{\perim{I(\poly)}} \leq \abs{\perim{\poly}} + c^*$, thus,
      $\abs{\perim{\poly'}} < \abs{\perim{\poly}}+c$, and
      since~$\poly'$ is a minimal-perimeter animal, we also know by the same premise
      that~$\abs{\perim{\poly'}} = \abs{\border{\poly'}}+c$, and, hence, 
      that~$\abs{\border{\poly'}} < \abs{\perim{\poly}}$.
      Consider now the animal~$D(\poly')$.
      Recall that $\abs{\poly'} = \abs{I(\poly)}=\abs{\poly}+\abs{\perim{\poly}}$, thus,
      the size of~$D(\poly')$ is at least~$\abs{\poly}+1$,
      and~$\abs{\perim{D(\poly')}} < \abs{\perim{\poly}} = \minp(n)$
      (since the perimeter of~$D(\poly')$ is a subset of the border of~$\poly'$).
      This is a contradiction to the first premise, which states that the
      sequence~$\minp(n)$ is monotone increasing.
      Hence, the animal~$\poly'$ cannot exist, and~$I(\poly)$ is a minimal-perimeter animal.
      \myqed
   \end{proof}

   We now proceed to demonstrating the effect of repeated inflation on the size of
   minimal-perimeter animals.

   \begin{lemma}
      \label{lemma:pnc_size}
      The minimum perimeter size of animals of size~$n+k\minp(n)+c^*k(k-1)/2$
      (for~$n > 1$ and any $k \in \N$) is~$\minp(n)+c^*k$.
   \end{lemma}

   \begin{proof}
      We repeatedly inflate a minimal-perimeter animal~$\poly$, whose initial size is~$n$.
      The size of the perimeter of~$\poly$ is~$\minp(n)$, thus, inflating
      it creates a new animal of size~$n+\minp(n)$, and the size of the border
      of~$I(\poly)$ is~$\minp(n)$, thus, the size of~$I(\poly)$
      is~$\minp(n) + c^*$.
      Continuing the inflation of the animal, the $k$th inflation will increase the size of the
      animal by $\minp(n) + (k-1)c^*$ and will increase the size of the perimeter by~$c^*$.
      Summing up these quantities yields the claim.
      \myqed
   \end{proof}

   Next, we prove that inflation preserves difference, that is, inflating two different
   minimal-perimeter animals (of equal or different sizes) always produces two different
   new animals.  (Note that this is not true for non-minimal-perimeter animals.)

   \begin{lemma}
      \label{lemma:different_inflating}
      Let~$\poly_1,\poly_2$ be two different minimal-perimeter animals.
      Then, regardless of whether or not~$\poly_1,\poly_2$ have the same size,
      the animals~$I(\poly_1)$ and~$I(\poly_2)$ are different as well.
   \end{lemma}

   \begin{proof}
      Assume to the contrary that $\poly = I(\poly_1) = I(\poly_2)$, that is,
      $\poly = \poly_1 \cup \perim{\poly_1} = \poly_2 \cup \perim{\poly_2}$.
      In addition, since $\poly_1 \neq \poly_2$, and since a cell cannot belong
      simultaneously to both an animal and to its perimeter, this means
      that~$\perim{\poly_1} \neq \perim{\poly_2}$.  The border of~$\poly$ is a
      subset of both~$\perim{\poly_1}$ and~$\perim{\poly_2}$, that is,
      $\border{\poly} \subset \perim{\poly_1} \cap \perim{\poly_2}$.
      Since~$\perim{\poly_1} \neq \perim{\poly_2}$, we obtain that
      either~$\abs{\border{\poly}} < \abs{\perim{\poly_1}}$
      or~$\abs{\border{\poly}} < \abs{\perim{\poly_2}}$;
      assume without loss of generality the former case.
      Now consider the animal~$D(\poly)$.
      Its size is~$\abs{\poly}-\abs{\border{\poly}}$.
      The size of~$\poly$ is~$\abs{\poly_1}+\abs{\perim{\poly_1}}$, thus,
      $\abs{D(\poly)} > \abs{\poly_1}$, and since the perimeter of~$D(\poly)$
      is a subset of the border of~$\poly$, we conclude
      that~$\abs{\perim{D(\poly)}} < \abs{\perim{\poly_1}}$.
      However, $\poly_1$ is a minimal-perimeter animal, which is a
      contradiction to the first premise of the theorem, which states
      that~$\minp(n)$ is monotone increasing.
      \myqed
   \end{proof}

   To complete the cycle, we also prove that for any minimal-perimeter
   animal~$\poly \in M_{n+\minp(n)}$, there is a minimal-perimeter source
   in~$M_n$, that is, an animal~$\poly'$ whose inflation yields~$\poly$.
   Specifically, this animal is $D(\poly)$.

   \begin{lemma}
      \label{lemma:deflating}
      For any~$\poly \in M_{n+\minp(n)}$, we also have
      that~$I(D(\poly)) = \poly$.
   \end{lemma}

   \begin{proof}
      Since~$\poly \in M_{n+\minp(n)}$, we have by
      Lemma~\ref{lemma:pnc_size} that~$\abs{\perim{\poly}} = \minp(n)+c^*$.
      Combining this with the equality~$\abs{\perim{\poly}} = \abs{\border{\poly}}+c^*$, we
      obtain that~$\abs{\border{\poly}} = \minp(n)$, thus, $\abs{D(\poly)} = n$
      and $\abs{\perim{D(\poly)}} \geq \minp(n)$.
      Since the perimeter of~$D(\poly)$ is a subset of the border of~$\poly$,
      and~$\abs{\border{\poly}} = \minp(n)$, we conclude that the perimeter
      of~$D(\poly)$ and the border of~$\poly$ are the same set of cells,
      and, hence, $I(D(\poly)) = \poly$.
      \myqed
   \end{proof}

   Let us now wrap up the proof of the main theorem.
   In Lemma~\ref{lemma:minimal-inflating} we have shown that for any
   minimal-perimeter animal~$\poly \in M_n$, we have that~$I(\poly) \in M_{n+\minp(n)}$.
   In addition, Lemma~\ref{lemma:different_inflating} states that the inflation of two different
   minimal-perimeter animals results in two other different minimal-perimeter animals.
   Combining the two lemmata, we obtain that~$\abs{M_n} \leq \abs{M_{n+\minp(n)}}$.
   On the other hand, in Lemma~\ref{lemma:deflating} we have shown that
   if~$\poly \in M_{n+\minp(n)}$, then~$I(D(\poly)) = \poly$, and, thus, for any animal
   in~$M_{n+\minp(n)}$, there is a unique source in~$M_n$ (specifically, $D(\poly)$),
   whose inflation yields~$\poly$.  Hence, $\abs{M_n} \geq \abs{M_{n+\minp(n)}}$.
   Combining the two relations, we conclude that~$\abs{M_n} = \abs{M_{n+\minp(n)}}$.
   \myqed
\end{proof} 

\subsection{Inflation Chains}

Theorem~\ref{thm:main} implies that there exist infinite chains of sets of minimal-perimeter
animals, each set obtained by inflating all members of the previous set, while the
cardinalities of all sets in a chain are equal.  Obviously, there are sets of
minimal-perimeter animals that are not created by the inflation of any other sets.
We call the size of animals in such sets an \emph{inflation-chain root}.
Using the definitions and proofs in the previous section, we are able to characterize which
sizes can be inflation-chain roots.
Then, using one more condition, which holds in the lattices
we consider, we determine which values are the actual inflation-chain roots.
To this aim, we define the pseudo-inverse function
\[
   \minp^{-1}(p) = \min\set{n \in \N \mid \minp(n) = p}.
\]
Since~$\minp(n)$ is a monotone-increasing discrete function, it is a step function,
and the value of~$\minp^{-1}(p)$ is the first point in each step.

\begin{theorem}
   \label{thm:root-candidates}
   Let~$\lattice$ be a lattice satisfying the premises of Theorem~\ref{thm:main}.
   Then, all inflation-chain roots are either~$\minp^{-1}(p)$
   or~$\minp^{-1}(p)-1$, for some $p \in \N$.
\end{theorem}

\begin{proof}
   Recall that~$\minp(n)$ is a step function, where each step represents all
   animal sizes for which the minimal perimeter is~$p$.
   Let us denote the start and end of the step representing the perimeter~$p$ by~$n_b^p$
   and~$n_e^p$, respectively.  Formally, $n_b^p = \minp^{-1}(p)$
   and~$n_e^p = \minp^{-1}(p+1)-1$.

   For each size~$n$ of animals in the step~$\bra{n_b^p,n_e^p}$, inflating a minimal-perimeter
   animal of size~$n$ results in an animal of size~$n{+}p$, and
   by~Lemma~\ref{lemma:pnc_size}, the perimeter of the inflated animal is~$p{+}c^*$.
   Thus, the inflation of animals of all sizes in the step of perimeter~$p$ yields animals
   that appear in the step of perimeter~$p{+}c^*$.
   In addition, they appear in a
   \emph{consecutive} portion of the step, specifically, the range~$\bra{n_b^p+p,n_e^p+p}$.
   Similarly, the step~$\bra{n_b^{p+1},n_e^{p+1}}$ is mapped by inflation to the
   range~$\bra{n_b^{p+1}+p+1,n_e^{p+1}+p+1}$, which is a portion of the step of~$p{+}1$.
   Note that the former range ends at~$n_e^p+p = n_b^{p+1}+p-1$, while the latter range
   starts at~$n_b^{p+1}+p+1$, thus, there is exactly one size of animals,
   specifically,~$n_b^{p+1}+p$, which is not covered by inflating animals in the
   ranges~$\bra{n_b^p+p,n_e^p+p}$ and~$\bra{n_b^{p+1},n_e^{p+1}}$.
   These two ranges represent two different perimeter sizes.  Hence, the size~$n_b^{p+1}+p$
   must be either the end of the first step, $n_e^{p+c^*}$,
   or the beginning of the second step,
   $n_b^{p+c^*+1}$. This concludes the proof.
   \myqed
\end{proof}

The arguments of the proof of Theorem~\ref{thm:root-candidates} are visualized in
Figure~\ref{fig:minpH_roots} for the case of polyhexes.
In fact, as we show below (see Theorem~\ref{thm:root-conditioned}), only the second
option exists, but in order to prove this, we also need a maximality-conservation
property of the inflation operation.

Here is another perspective for the above result.
Note that minimal-perimeter animals, with size corresponding to $n_e^{p}$ (for
some~$p \in \N$), are the largest animals with perimeter~$p$.
Intuitively, animals with the largest size, for a certain perimeter size, tend to be
``spherical'' (``round'' in two dimensions), and inflating them makes them even more spherical.
Therefore, one might expect that for a general lattice, the inflation operation will preserve
the property of animals being the largest for a given perimeter.  In fact, this has been proven
rigorously for the square lattice~\cite{altshuler2006,sieben2008polyominoes} and for the
hexagonal lattice~\cite{VainsencherB08,fulep2010polyiamonds}.
However, this also means that inflating a minimal-perimeter animal of size~$n_e^p$
yields a minimal-perimeter animal of size~$n_e^{p+c^*}$, and, thus, $n_e^p$ cannot be an
inflation-chain root.  We summarize this discussion in the following theorem.

\begin{theorem}
   \label{thm:root-conditioned}
   Let~$\lattice$ be a lattice for which the three premises of Theorem~\ref{thm:main} are
   satisfied, and, in addition, the following condition holds.
   \begin{enumerate}[label=(\arabic*)]
       \setcounter{enumi}{3}
       \item The inflation operation preserves the property of having a maximum size for a
             given perimeter.
   \end{enumerate}
   Then, the inflation-chain roots are precisely~$(\minp_\lattice)^{-1}(p)$, for all $p \in \N$.
   \myqed
\end{theorem}

\subsection{Convergence of Inflation Chains}

We now discuss the structure of inflated animals, and show that 
under a certain condition, inflating repeatedly \emph{any} animal (or
actually, any set, possibly disconnected, of lattice cells) ends up in a 
minimal-perimeter animal after a finite number of inflation steps.

Let~$I^k(Q)$ ($k>0$) denote the result of applying repeatedly~$k$ times the inflating
operator~$I(\cdot)$, starting from the animal~$Q$.  Equivalently, 
\[
   I^k(Q) = Q \cup \set{c \mid \mbox{Dist}(c,Q) \leq k},
\]
where~$\mbox{Dist}(c,Q)$ is the Lattice distance from a cell~$c$ to the animal~$Q$.
For brevity, we will use the notation $Q^k = I^k(Q)$.

Let us define the function 
\(
   \phi(Q) = \minn(\abs{\perim{Q}}) - \abs{Q}
\)
and explain its meaning.
When~$\phi(Q) \geq 0$, it counts the cells that should be added to~$Q$, with no change
to its perimeter, in order to make it a minimal-perimeter animal.
In particular, if~$\phi(Q) = 0$, then~$Q$ is a minimal-perimeter animal.
Otherwise, if~$\phi(Q) < 0$, then~$Q$ is also a minimal-perimeter animal, and $\abs{\phi(Q)}$ cells can be
removed from~$Q$ while still keeping the result a minimal-perimeter animal and without changing its perimeter.

\begin{lemma}
   \label{lemma:jumps-p-1}
   For any value of~$p$, we have that~$\minn(p+c^*)-\minn(p) = p-1$.
\end{lemma}

\begin{proof}
   Let~$Q$ be a minimal-perimeter animal with area~$n_b^p = \minn(p)$.
   The area of~$I(Q)$ is~$n_b^p+p$, thus, by Theorem~\ref{thm:main},
   $\perim{I(Q)} = p+c^*$.  The area~$n_b^{p+c^*}$ is an
   inflation-chain root, hence, the area of~$I(Q)$ cannot be~$n_b^{p+c^*}$.
   Except~$n_b^{p+c^*}$, animals of all other areas in the
   range~$[n_b^{p+c^*},\dots,n_e^{p+c^*}]$ are created by inflating
   minimal-perimeter animals with perimeter~$p$.
   The animal~$Q$ is of area~$n_b^p$, \emph{i.e.}, the area of~$I(Q)$ must
   be the minimal area from $\bra{n_b^{p+c^*},n_e^{p+c^*}}$ which is not
   an inflation-chain root.  Hence, the area of~$I(Q)$ is~$n_b^{p+c^*}+1$.
   We now equate the two expressions for the area of $I(Q)$:
   $n_b^p+p = n_b^{p+c^*}+1$.  That is, $n_b^{p+c^*}-n_b^{p} = p-1$.
   The claim follows.
\end{proof}

Using Lemma~\ref{lemma:jumps-p-1}, we can deduce the following result.

\begin{lemma}
   \label{lem:conv-step}
   If~$\abs{\perim{I(Q)}} = \abs{\perim{Q}} +c^*$,
   then~$\phi(I(Q)) = \phi(Q)-1$.
\end{lemma}

\begin{proof}
   \begin{align*}
      \phi(I(Q)) &= \minn(\abs{\perim{I(Q)}}) - \abs{I(Q)} \\
         &= \minn(\abs{\perim{Q}}+c^*) - (\abs{Q} + \abs{\perim{Q}}) \\
         &= \minn(\abs{\perim{Q}}) + \abs{\perim{Q}} -1 - \abs{Q}
               - \abs{\perim{Q}} \\
         &= \minn(\abs{\perim{Q}}) -\abs{Q} - 1 \\
         &= \phi(Q) - 1.
   \end{align*}
\end{proof}

Lemma~\ref{lem:conv-step} tells us that inflating an animal, $Q$, which satisfies
$\abs{\perim{I(Q)}} = \abs{\perim{Q}} +c^*$, reduces $\phi(Q)$ by $1$.
In other words, $I(Q)$ is ``closer'' than~$Q$ to being a minimal-perimeter animal.
This result is stated more formally in the following theorem.

\begin{theorem}
   \label{thm:convergence}
   Let~$\lattice$ be a lattice for which the four premises of Theorems~\ref{thm:main}
   and~\ref{thm:root-conditioned} are satisfied, and, in addition, the following condition holds.
   \begin{enumerate}[label=(\arabic*)]
      \setcounter{enumi}{4}
      \item For every animal~$Q$, there exists some finite number~$k_0 = k_0(Q)$, such that for
            every $k>k_0$, we have that~$\abs{\perim{Q^{k+1}}} = \abs{\perim{Q^{k}}} + c$.
   \end{enumerate}
   Then, after a finite number of inflation steps, 
   any animal becomes a minimal-perimeter animal.
\end{theorem}

\begin{proof}
   The claim follows from Lemma~\ref{lem:conv-step}.
   After~$k_0$ inflation operations, the premise of this lemma holds.
   Then, any additional inflation step will reduce~$\phi(Q)$ by~$1$ until~$\phi(Q)$ is nullified, which is
   precisely when the animal becomes a minimal-perimeter animal.
   (Any additional inflation steps would add superfluous cells, in the sense that they can be removed while keeping
   the animal a minimal-perimeter animal.)
\end{proof}


\section{Polyominoes}
\label{sec:polyominoes}

Throughout this section, we consider the two-dimensional square lattice~$\squ$, and
show that the premises of Theorem~\ref{thm:main} hold for this lattice.
The lattice-specific notation ($M_n$, $\minp(n)$, and~$c^*$) in this section refer to~$\squ$.

\subsection{Premise 1:  Monotonicity}

The function~$\minp^\squ(n)$, that gives the minimum possible size of the perimeter of a
polyomino of size~$n$, is known to be weakly-monotone increasing.
This fact was proved independently by Altshuler et al.~\cite{altshuler2006} and by
Sieben~\cite{sieben2008polyominoes}.
The latter reference also provides the following explicit formula.

\begin{theorem}
   \label{thm:minp_sqr}
   \textup{\cite[Thm.~$5.3$]{sieben2008polyominoes}}
   $\minp^\squ(n) = \ceil{\sqrt{8n-4} \,}+2$.
   \myqed 
\end{theorem}

\subsection{Premise 2:  Constant Inflation}

The second premise is apparently the hardest to show.
We will prove that it holds for~$\squ$ by analyzing the patterns which may appear on the
border of minimal-perimeter polyominoes.

Asinowski et al.~\cite{asinowski2017enumerating} defined the \emph{excess} of a perimeter cell
as the number of adjacent occupied cell minus one, and the total \emph{perimeter excess} of an
animal~$\poly$, $e_P(\poly)$, as the sum of excesses over all perimeter cells of~$\poly$.
We extend this definition to border cells, and, in a similar manner, define the \emph{excess} of
a border cell as the number of adjacent empty cells minus one, and the \emph{border excess}
of~$\poly$, $e_B(\poly)$, as the sum of excesses over all border cells of~$\poly$.

First, we establish a connection between the size of the perimeter of a polyomino to the size
of its border.
The following formula is universal for all lattice animals.

\begin{lemma}
   \label{lemma:pebe}
   For every animal~$\poly$, we have that
   \begin{equation}
      \label{eq:pebe}
      \abs{\perim{\poly}} + e_P(\poly) = \abs{\border{\poly}} + e_B(\poly).
   \end{equation}
\end{lemma}

\begin{proof}
   Consider the (one or more) rectilinear polygons bounding the animal~$\poly$.
   The two sides of the equation are equal to the total length of the polygon(s) in terms of
   lattice edges.
   Indeed, this length can be computed by iterating over either the border or the perimeter
   cells of~$\poly$.  In both cases, each cell contributes one edge plus its excess to the
   total length.  The claim follows.
   \myqed
\end{proof}

\begin{figure}
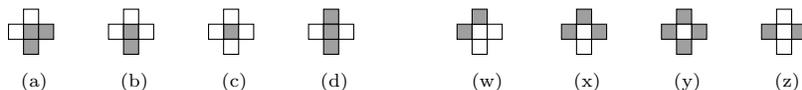

   \centering
   \begin{subfigure}[t]{0.1\textwidth}
      \centering
      \drawpoly[scale=0.5]{bs1.txt}
      \caption{}
   \end{subfigure}
   \begin{subfigure}[t]{0.1\textwidth}
      \centering
      \drawpoly[scale=0.5]{bs2.txt}
      \caption{}
   \end{subfigure}
   \begin{subfigure}[t]{0.1\textwidth}
      \centering
      \drawpoly[scale=0.5]{bs3.txt}
      \caption{}
   \end{subfigure}
   \begin{subfigure}[t]{0.1\textwidth}
      \centering
      \drawpoly[scale=0.5]{bs4.txt}
      \caption{}
   \end{subfigure}
   \qquad
   \begin{subfigure}[t]{0.1\textwidth}
      \centering
      \drawpoly[scale=0.5]{ps1.txt}
      \addtocounter{subfigure}{18} 
      \caption{}
   \end{subfigure}
   \begin{subfigure}[t]{0.1\textwidth}
      \centering
      \drawpoly[scale=0.5]{ps2.txt}
      \caption{}
   \end{subfigure}
   \begin{subfigure}[t]{0.1\textwidth}
      \centering
      \drawpoly[scale=0.5]{ps3.txt}
      \caption{}
   \end{subfigure}
   \begin{subfigure}[t]{0.1\textwidth}
      \centering
      \drawpoly[scale=0.5]{ps4.txt}
      \caption{}
   \end{subfigure}
   \caption{All possible patterns of cells, up to symmetries, with positive excess.
            The gray cells are polyomino cells, while the white cells are perimeter cells.
            The centers of the ``crosses'' are the subject cells, and the patterns show
            the immediate neighbors of these cells.
            Patterns~(a--d) exhibit excess border cells, while Patterns~(w--z) exhibit
            excess perimeter cells.
            }
   \label{fig:patterns_sqr}
\end{figure}

\begin{figure}
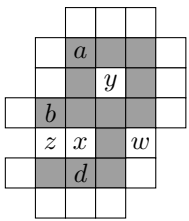

    \centering
   \drawpoly{patterns.txt}
    \caption{A~sample polyomino with marked patterns.}
    \label{fig:patterns-exmp}
\end{figure}

Let~$\#\square$ be the number of excess cells of a certain type in a polyomino,
where~`$\square$' is one of the symbols~$a$--$d$ and~$w$--$z$,
as classified in Figure~\ref{fig:patterns_sqr}.
Figure~\ref{fig:patterns-exmp} depicts a polyomino which includes cells of all these types.
Counting~$e_P(Q)$ and~$e_B(Q)$ as functions of the different patterns of
excess cells, we see that 
\(
   e_B(Q) = \#a + 2\#b + 3\#c + \#d
\)
and
\(
   e_P(Q) = \#w + 2\#x + 3\#y + \#z.
\)
Substituting~$e_B$ and~$e_P$ in \myeqref{eq:pebe}, we obtain that
\[
   \psize = \bsize +  \#a + 2\#b+3\#c + \#d - \#w - 2\#x-3\#y - \#z.
\]
Since Pattern~(c) is a singleton cell, we can ignore it in the general
formula. Thus, we have that
\[
   \psize = \bsize +  \#a + 2\#b + \#d - \#w - 2\#x-3\#y - \#z.
\]

We now simplify the equation above, first by eliminating the hole pattern, namely, Pattern~(y).

\begin{lemma}
   \label{lemma:no-holes-sqr}
   Any minimal-perimeter polyomino is simply connected (that is, it
   does not contain holes).
\end{lemma}

\begin{proof}
   The sequence~$\minp(n)$ is weakly-monotone increasing.\footnote{
      In the sequel, we simply say ``monotone increasing.''
   }
   Assume that there exists a minimal-perimeter polyomino~$\poly$ with a
   hole. Consider the polyomino~$\poly'$ that is obtained by filling this
   hole. The area of~$\poly'$ is clearly larger than that of~$\poly$,
   however, the perimeter size of~$\poly'$ is smaller than that of~$\poly$ since we eliminated
   the perimeter cells inside the hole but did not introduce new perimeter cells.
   This is a contradiction to~$\minp(n)$ being monotone increasing.
   \myqed
\end{proof}

Next, we continue to eliminate terms from the equation by showing some invariant related to the
turns of the boundary of a minimal-perimeter polyomino.

\begin{lemma}
   \label{lemma:sum_of_turns}
   For a simply connected polyomino, we have that
   \(
      \#a +2\#b -\#w -2\#x = 4.
   \)
\end{lemma}

\begin{proof}
   The boundary of a polyomino without holes is a simple polygon, thus,
   the sum of its internal angles is $(v-2)\pi$,
   where~$v$ is the complexity (number of vertices) of the polygon.
   Note that Pattern~(a) (resp.,~(b)) adds one (resp., two)
   $\pi/2$-vertex to the polygon.
   Similarly, Pattern~(w) (resp.~(x)) adds one (resp., two) $3\pi/2$-vertex.
   All other patterns do not involve vertices.
   Let~$L = \#a+2\#b$ and~$R =  \#w+2\#x$.
   Then, the sum of angles of the boundary polygon implies that
   $L \cdot \pi/2 + R \cdot 3\pi/2 = (L+R-2) \cdot \pi$,
   that is, $L-R = 4$. The claim follows.
   \myqed
\end{proof}
 
Finally, we show that Patterns~(d) and~(z) cannot exist in a minimal-perimeter polyomino.

We define a \emph{bridge} as a cell whose removal renders the polyomino disconnected.
Similarly, a perimeter bridge is a perimeter cell that neighbors two or more
connected components of the complement of the polyomino.
Observe that minimal-perimeter polyominoes do not contain any bridges, \emph{i.e.},
cells of Patterns~(d) or~(z).  This is stated in the following lemma.

\begin{lemma}
   \label{lemma:no-bridges-sqr}
   A minimal-perimeter polyomino does not contain any bridge cells.
\end{lemma}

\begin{proof}
     Let~$\poly$ be a minimal-perimeter polyomino. For the sake of
   contradiction, assume first that there is a cell~$f \in \perim{\poly}$
   as part of Pattern~(z). Assume without loss of generality
   that the two adjacent polyomino cells are to the left and to
   the right of~$f$. These two cells must be connected, thus, the area
   below (or above)~$f$ must form a cavity in the polyomino shape.
   Let, then, $\poly'$ obtained by adding~$f$ to~$\poly$ and filling the cavity.
   \figrefs{fig:no_z+d}(a,b) illustrate this situation.
   The cell directly above~$f$ becomes a perimeter cell, the cell~$f$
   ceases to be a perimeter cell, and at least one perimeter cell in the
   area filled below~$f$ is eliminated,
   thus,~$\abs{\perim{\poly'}} < \abs{\perim{\poly}}$
   and~$\abs{\poly'} > \abs{\poly}$,
   which is a contradiction to the sequence~$\minp(n)$ being
   monotone increasing.
   Therefore, polyomino~$\poly$ does not contain perimeter cells that
   fit Pattern~(z).
   
   \begin{figure}
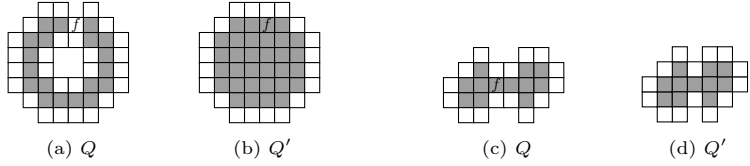

      \centering
      \begin{subfigure}[t]{0.2\textwidth}
         \centering
         \drawpoly[scale=0.5, every node/.style={scale=0.6}]{no_z1.txt}
         \caption{$\poly$}
      \end{subfigure}
      \begin{subfigure}[t]{0.2\textwidth}
         \centering
         \drawpoly[scale=0.5, every node/.style={scale=0.6}]{no_z2.txt}
         \caption{$\poly'$}
      \end{subfigure}
      \qquad
      \begin{subfigure}[t]{0.2\textwidth}
         \centering
         \drawpoly[scale=0.5, every node/.style={scale=0.6}]{no_d1.txt}
         \caption{$\poly$}
      \end{subfigure}
      \begin{subfigure}[t]{0.2\textwidth}
         \centering
         \drawpoly[scale=0.5]{no_d2.txt}
         \caption{$\poly'$}
      \end{subfigure}
      \caption{Forbidden patterns for the proof of Theorem~\ref{theorem:pb4}.}
      \label{fig:no_z+d}
   \end{figure}

   Now assume for contradiction that~$\poly$ contains a cell~$f$ that forms
   Pattern~(d).  Let~$\poly'$ be the polyomino obtained from~$\poly$ by
   removing~$f$ (this will break~$\poly'$ into two separate pieces) and then
   shifting to the left the piece on the right (this will unite the two pieces
   into a new polyomino).
   \figrefs{fig:no_z+d}(c,d) demonstrate this situation.
   This operation is always valid since~$\poly$ is of minimal perimeter,
   hence, by Lemma~\ref{lemma:no-holes-sqr}, it is simply connected, and thus,
   removing~$f$ breaks~$\poly$ into two separate polyominoes with a gap of one
   cell in between.  Shifting to the left the piece on the right will not create a
   collision since this would mean that the two pieces were touching, which is not the case.
   On the other hand, the shift will eliminate the gap that was created by the
   removal of~$f$, hence, the two pieces will now form a new connected polyomino.
   The area of~$\poly'$
   is one less than the area of~$\poly$, and the perimeter of~$\poly'$ is
   smaller by at least two than the perimeter of~$\poly$, since the
   perimeter cells below and above~$f$ cease to be part of the perimeter,
   and connecting the two parts does not create new perimeter cells.
   From the formula of~$\minp(n)$, we know that
   $\minp(n)-\minp(n-1) \leq 1$ for~$n \geq 3$.
   However, $\abs{\poly} - \abs{\poly'} = 1$
   and~$\abs{\perim{\poly}} - \abs{\perim{\poly'}} = 2$, hence,
   $\poly$ is not a minimal-perimeter polyomino, which contradicts our
   assumption.
   Therefore, there are no cells in~$\poly$ that fit Pattern~(d).
   This completes the proof. \myqed
\end{proof}

We are now ready to wrap up the proof of the constant-inflation theorem.

\begin{theorem}
   \label{theorem:pb4}
   (Stepping Theorem)
   For any minimal-perimeter polyomino~$\poly$ (except the singleton cell), we have
   that $\psize=\bsize+4.$
\end{theorem}

\begin{proof}
   Lemma~\ref{lemma:sum_of_turns} tells us that~$\psize=\bsize+4+\#d-\#z$.
   By Lemma~\ref{lemma:no-bridges-sqr}, we know that $\#d = \#z = 0$.
   The claim follows at once.
   \myqed
\end{proof}

\subsection{Premise 3:  Deflation Resistance}

\begin{lemma}
   \label{lemma:def_valid}
   Let~$\poly$ be a minimal-perimeter polyomino of area~$n+\minp(n)$
   (for $n \geq 3$). Then, $D(\poly)$ is a valid (connected) polyomino.
\end{lemma}

\begin{proof}
   Assume to the contrary that~$D(\poly)$ is not connected, so that it is
   composed of at least two connected parts.
   Assume first that~$D(\poly)$ is composed of exactly two parts,
   $\poly_1$ and~$\poly_2$.
   Define the \emph{joint perimeter} of the two parts,
   $\perim{\poly_1,\poly_2}$, to be~$\perim{\poly_1} \cup \perim{\poly_2}$.
   Since~$\poly$ is a minimal-perimeter polyomino of area $n+\minp(n)$,
   we know by Theorem~\ref{theorem:pb4}
   that its perimeter size is~$\minp(n)+4$ and its
   border size is~$\minp(n)$, respectively.
   Thus, the size of~$D(\poly)$ is exactly~$n$ regardless of whether or
   not~$D(\poly)$ is connected.
   Since deflating~$\poly$ results in~$\poly_1 \cup \poly_2$,
   the polyomino~$\poly$ must have an (either horizontal, vertical, or
   diagonal) ``bridge'' of border cells which disappears by the deflation.
   The width of the bridge is at most~2, thus,
   $\abs{\perim{\poly_1} \cap \perim{\poly_2}} \leq 2$. Hence,
   $\abs{\perim{\poly_1}} + \abs{\perim{\poly_2}} - 2 \leq
       \abs{\perim{\poly_1,\poly_2}}$. 
   Since~$\perim{\poly_1,\poly_2}$ is a subset of~$\border{\poly}$,
   we have that $\abs{\perim{\poly_1,\poly_2}} \leq \minp(n)$. Therefore,
   \begin{equation}
      \label{eq:def_valid_1}
      \minp(\abs{\poly_1}) + \minp(\abs{\poly_2}) - 2 \leq \minp(n). 
   \end{equation}

   Recall that~$\abs{\poly_1} + \abs{\poly_2} = n$.
   It is easy to observe that~$\minp(\abs{\poly_1})+\minp(\abs{\poly_2})$
   is minimized when~$\abs{\poly_1}=1$ and $\abs{\poly_2} = n-1$ (or vice
   versa).  Had the function~$\minp(n)$ (shown in \figref{fig:minp_plot})
   \begin{figure}
      \centering
      \includegraphics[scale=0.4]{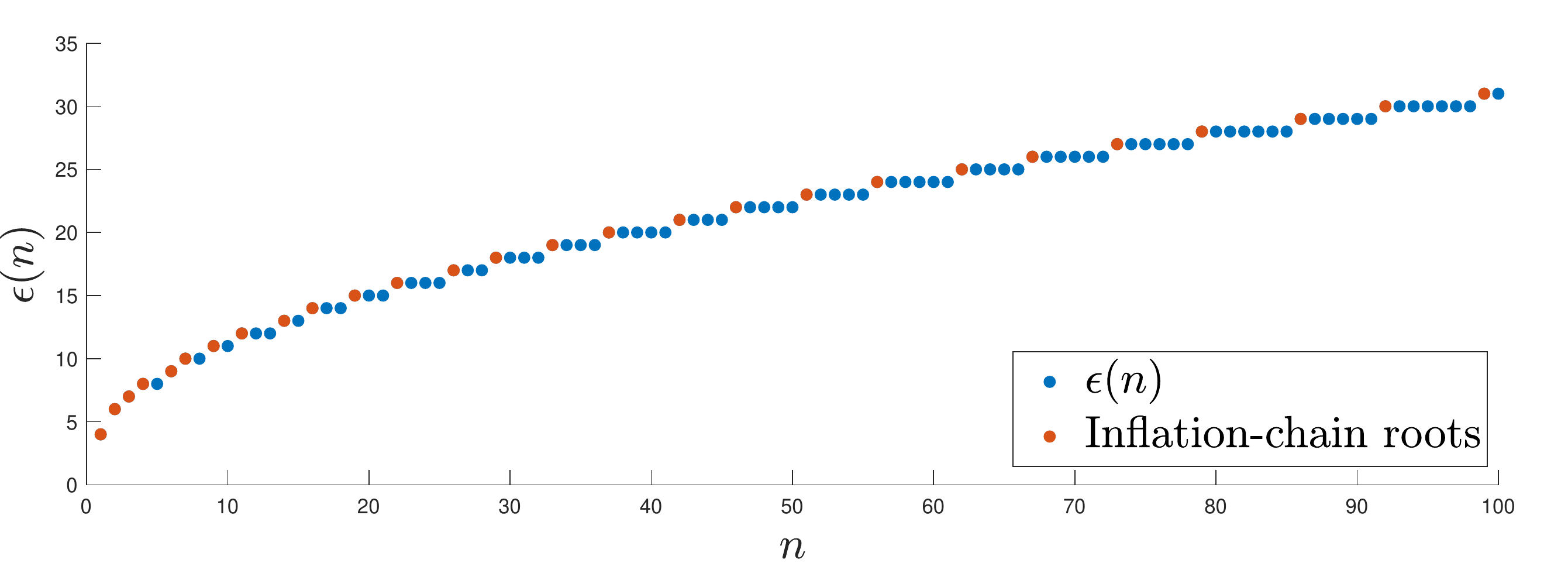}
      \caption{The function~$\minp(n)$.}
      \label{fig:minp_plot}
   \end{figure}
   been $2+\sqrt{8n-4}$ (without rounding up), this would be obvious.
   But since $\minp(n) = \left\lceil 2+\sqrt{8n-4} \, \right\rceil$, it is a
   step function (with an infinite number of intervals), where the gap
   between all successive steps is exactly~1, except the gap between the
   two leftmost steps which is~2.  This guarantees that despite the
   rounding, the minimum of~$\minp(\abs{\poly_1})+\minp(\abs{\poly_2})$
   occurs as claimed.
   Substituting this into \myeqref{eq:def_valid_1}, and using the
   fact that~$\minp(1)=4$, we see that $\minp(n-1) + 2 \leq \minp(n)$.
   However, we know~\cite{sieben2008polyominoes} that
   $\minp(n) - \minp(n-1) \leq 1$ for $n\geq 3$, which is a contradiction.
   Thus, the deflated version of~$\poly$ cannot split into two parts unless it splits into two
   singleton cells, which is indeed the case for a minimal-perimeter
   polyomino of size~8, specifically,
   \(
      D( \!\! \raisebox{-2.25mm}{\drawpoly[scale=0.3]{diag8.txt}} \!\! ) = \!\!
         \raisebox{-1.5mm}{\drawpoly[scale=0.3]{diag2.txt}}
   \).

   The same method can be used for showing that~$D(\poly)$ cannot be composed
   of more then two parts.  Note that this proof does not hold for
   polyominoes of area which is not of the form~$n+\minp(n)$, but it
   suffices for the use in Theorem~\ref{thm:main}.
   \myqed
\end{proof}

As mentioned earlier, it was already proven elsewhere~\cite{altshuler2006,sieben2008polyominoes}
that Premise~4 (roots of inflation chains) is fulfilled for the square lattice.
Therefore, we proceed to showing that Premise~5 holds.

\subsection{Premise 5:  Convergence to a Minimum-Perimeter Polyomino}

In this section, we show that starting from any polyomino~$P$, and applying repeatedly some finite number
of inflation steps, we obtain a polyomino $Q = Q(P)$, for which $\perim{I(Q)} = \perim{Q} + 4$.
Let~$R(Q)$ denote the \emph{diameter} of~$Q$, \emph{i.e.}, the maximal horizontal or
vertical distance ($L^\infty$) between two cells of~$Q$.
The following lemma shows that some geometric features of a polyomino
disappear after inflating it enough times.

\begin{lemma}
   \label{lem:no-hdz}
   For any~$k > R(Q)$, the polyomino~$Q^k$ does not contain any
   (i)~holes; (ii)~cells of Type~(d); or (iii)~patterns of Type~(z).
\end{lemma}
\begin{proof}

   \begin{itemize}
   \item[(i)]
   Let~$Q$ be a polyomino, and assume that~$Q^k$ contains a hole.
   Consider a cell~$c$ inside the hole, and let~$c_u$ be the cell 
   of~$Q^k$ that lies immediately above it.  (Note that since~$c_u$
   belongs to the border of~$Q^k$, it is not a cell of~$Q$.)  Any cell
   that resides (not necessarily directly) below~$c$ is closer to~$c$ than
   to~$c_u$.  Since $c_u \in Q^k$, it ($c_u$) is closer than~$c$ to~$Q$, thus,
   there must be a cell of~$Q$ (not necessarily directly) above~$c$,
   otherwise~$c_u$ would not belong to~$Q^k$.
   The same holds for cells below, to the right, and to the left
   of~$c$, thus,~$c$ resides within the axis-aligned bounding box of
   the extreme cells of~$Q$, and after~$R(Q)$ steps,~$c$ will be
   occupied, and any hole will be eliminated.

   \item[(ii)]
   Assume that there exists a polyomino~$Q$, for which the polyomino~$Q^k$ contains a
   cell of Type~(d).
   Without loss of generality, assume that the neighbors of~$c$ reside
   to its left and to its right, and denote them by $c_\ell,c_r$, respectively.   
   Denote by~$c_o$ one of the cells whose inflation created~$c_\ell$, \emph{i.e.},
   a cell which belongs to~$Q$ and is in distance of at most~$k$ from~$c_\ell$. 
   In addition, denote by $c_u,c_d$ the adjacent perimeter cells which
   lie immediately above and below~$c$, respectively.  The cell~$c_d$ is not
   occupied, thus, its distance from~$c_o$ is~$k+1$, which means
   that~$c_o$ lies in the same row as~$c_\ell$.  Assume for contradiction
   that~$c_o$ lies in a row below~$c_\ell$.  Then, the distance between~$c_o$ and~$c_d$
   is at most~$k$, hence~$c_d$ belongs to~$Q^k$.
   The same holds for~$c_u$; thus, cell~$c_o$ must lie in the same row as~$c_\ell$.
   Similar considerations show that~$c_o$ must lie to the left of~$c_\ell$,
   otherwise~$c_d$ and~$c_u$ would be occupied.
   In the same manner, one of the cells that originated~$c_r$
   must lie in the same row as~$c_r$ on its right.
   Hence, any cell of Type~(d) have cells of~$Q$ to its right and to its left, 
   and thus, it is found inside the bounding axis-aligned bounding box of~$Q$,
   which will necessarily be filled with polyomino cells after~$R(Q)$ inflation
   steps.

   \item[(iii)]
   Let~$c$ be a Type-(z) perimeter cell of~$Q^k$.  Assume, without loss
   of generality, that the polyomino cells adjacent to it are to its
   left and to its right, and denote them by~$c_\ell$ and~$c_r$, respectively.
   Let~$c_o$ denoted a cell whose repeated inflation has added~$c_\ell$ to $Q^k$.
   (Note that~$c_o$ might not be unique.)
   This cell must lie to the left of~$c$, otherwise, it will be closer to~$c$ than to~$c_\ell$,
   and~$c$ would not be a perimeter cell.
   In addition, $c_o$ must lie in the same row as~$c_\ell$, for otherwise, by the same 
   considerations as above, one of the cells above or below~$c$ will be occupied.
   The same holds for~$c_r$ (but to its right), thus, cells of Type~(z) must
   reside between two original cells of~$Q$, \emph{i.e.}, inside the bounding box
   of~$Q$, and after~$R(Q)$ inflation steps, all cells inside this box
   will become polyomino cells.
   \end{itemize}
\end{proof}

We can now conclude that inflating a polyomino~$Q$ for~$R(Q)$ times eliminates all
holes and bridges, and, thus, the polyomino~$Q^k$ will obey the equation
$\abs{\perim{Q^k}} = \abs{\border{Q^k}} + 4$.

\begin{lemma}
   \label{lem:conv-pb4}
   Let~$Q$ be a polyomino, and let~$k = R(Q)$.  We have that
   $\abs{\perim{Q^k}} = \abs{\border{Q^k}} + 4$.
\end{lemma}

\begin{proof}
   This follows at once from Lemma~\ref{lem:no-hdz} and
   Theorem~\ref{theorem:pb4}.
\end{proof}


\section{Polyhexes}
\label{sec:polyhexes}

In this section, we show that the premises of Theorem~\ref{thm:main} hold for the
two-dimensional hexagonal lattice~$\hex$.  The roadmap followed in this section is
similar to the one used in Section~\ref{sec:polyominoes}.  In this section, all the
lattice-specific notations refer to~$\hex$.

\subsection{Premise 1:  Monotonicity}

The first premise has been proven for~$\hex$ independently by Vainsencher and
Bruckstien~\cite{VainsencherB08} and by F\"{u}lep and Sieben~\cite{fulep2010polyiamonds}.
We will use the latter, stronger version which also includes a formula for $\minp(n)$.

\begin{theorem}
   \label{thm:minp_hex}
   \textup{\cite[Thm.~$5.12$]{fulep2010polyiamonds}}
   $\minp(n) = \ceil{\sqrt{12n-3}\,}+3$.
   \myqed 
\end{theorem}

Clearly, the function~$\minp(n)$ is weakly-monotone increasing.

\subsection{Premise 2:  Constant Inflation}

\begin{figure}
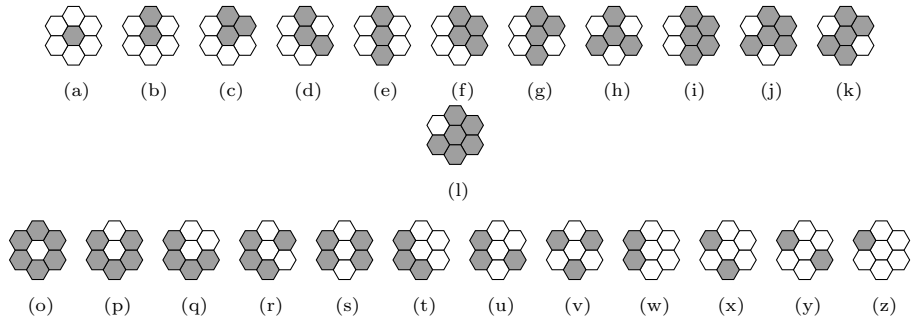

   \centering
   \begin{subfigure}[t]{0.075\textwidth}
      \centering
      \drawpolyhex[scale=0.65]{b0.txt}
      \caption{}
      \label{fig:b0}
   \end{subfigure}
   \begin{subfigure}[t]{0.075\textwidth}
      \centering
      \drawpolyhex[scale=0.65]{b1.txt}
      \caption{}
      \label{fig:b1}
   \end{subfigure}
   \begin{subfigure}[t]{0.075\textwidth}
      \centering
      \drawpolyhex[scale=0.65]{b2.txt}
      \caption{}
      \label{fig:b2}
   \end{subfigure}
   \begin{subfigure}[t]{0.075\textwidth}
      \centering
      \drawpolyhex[scale=0.65]{b3.txt}
      \caption{}
      \label{fig:b3}
   \end{subfigure}
   \begin{subfigure}[t]{0.075\textwidth}
      \centering
      \drawpolyhex[scale=0.65]{b4.txt}
      \caption{}
      \label{fig:b4}
   \end{subfigure}
   \begin{subfigure}[t]{0.075\textwidth}
      \centering
      \drawpolyhex[scale=0.65]{b5.txt}
      \caption{}
      \label{fig:b5}
   \end{subfigure}
   \begin{subfigure}[t]{0.075\textwidth}
      \centering
      \drawpolyhex[scale=0.65]{b6.txt}
      \caption{}
      \label{fig:b6}
   \end{subfigure}
   \begin{subfigure}[t]{0.075\textwidth}
      \centering
      \drawpolyhex[scale=0.65]{b7.txt}
      \caption{}
      \label{fig:b7}
   \end{subfigure}
   \begin{subfigure}[t]{0.075\textwidth}
      \centering
      \drawpolyhex[scale=0.65]{b8.txt}
      \caption{}
      \label{fig:b8}
   \end{subfigure}
   \begin{subfigure}[t]{0.075\textwidth}
      \centering
      \drawpolyhex[scale=0.65]{b9.txt}
      \caption{}
      \label{fig:b9}
   \end{subfigure}
   \begin{subfigure}[t]{0.075\textwidth}
      \centering
      \drawpolyhex[scale=0.65]{b10.txt}
      \caption{}
      \label{fig:b10}
   \end{subfigure}
   \begin{subfigure}[t]{0.075\textwidth}
      \centering
      \drawpolyhex[scale=0.65]{b11.txt}
      \caption{}
      \label{fig:b11}
   \end{subfigure}
   \medskip \\
   \begin{subfigure}[t]{0.075\textwidth}
      \centering
      \addtocounter{subfigure}{2} 
      \drawpolyhex[scale=0.65]{p0.txt}
      \caption{}
      \label{fig:p0}
   \end{subfigure}
   \begin{subfigure}[t]{0.075\textwidth}
      \centering
      \drawpolyhex[scale=0.65]{p1.txt}
      \caption{}
      \label{fig:p1}
   \end{subfigure}
   \begin{subfigure}[t]{0.075\textwidth}
      \centering
      \drawpolyhex[scale=0.65]{p2.txt}
      \caption{}
      \label{fig:p2}
   \end{subfigure}
   \begin{subfigure}[t]{0.075\textwidth}
      \centering
      \drawpolyhex[scale=0.65]{p3.txt}
      \caption{}
      \label{fig:p3}
   \end{subfigure}
   \begin{subfigure}[t]{0.075\textwidth}
      \centering
      \drawpolyhex[scale=0.65]{p4.txt}
      \caption{}
      \label{fig:p4}
   \end{subfigure}
   \begin{subfigure}[t]{0.075\textwidth}
      \centering
      \drawpolyhex[scale=0.65]{p5.txt}
      \caption{}
      \label{fig:p5}
   \end{subfigure}
   \begin{subfigure}[t]{0.075\textwidth}
      \centering
      \drawpolyhex[scale=0.65]{p6.txt}
      \caption{}
      \label{fig:p6}
   \end{subfigure}
   \begin{subfigure}[t]{0.075\textwidth}
      \centering
      \drawpolyhex[scale=0.65]{p7.txt}
      \caption{}
      \label{fig:p7}
   \end{subfigure}
   \begin{subfigure}[t]{0.075\textwidth}
      \centering
      \drawpolyhex[scale=0.65]{p8.txt}
      \caption{}
      \label{fig:p8}
   \end{subfigure}
   \begin{subfigure}[t]{0.075\textwidth}
      \centering
      \drawpolyhex[scale=0.65]{p9.txt}
      \caption{}
      \label{fig:p9}
   \end{subfigure}
   \begin{subfigure}[t]{0.075\textwidth}
      \centering
      \drawpolyhex[scale=0.65]{p10.txt}
      \caption{}
      \label{fig:p10}
   \end{subfigure}
   \begin{subfigure}[t]{0.075\textwidth}
      \centering
      \drawpolyhex[scale=0.65]{p11.txt}
      \caption{}
      \label{fig:p11}
   \end{subfigure}
   \caption{All possible patterns (up to symmetries) of border (first row) and perimeter
            (second row) cells.
            The gray cells are polyhex cells, while the white cells are perimeter cells.
            Each subfigure shows a cell in the middle, and the possible pattern of cells
            surrounding it.}
   \label{fig:patterns_hex}
\end{figure}

To show that the second premise holds, we analyze the different patterns that may
appear in the border and perimeter of minimal-perimeter polyhexes.
We can classify every border or perimeter cell by one of exactly~24 patterns,
distinguished by the number and positions of their adjacent occupied cells.
The~24 possible patterns are shown in Figure~\ref{fig:patterns_hex}.

Let us recall the equation subject of Lemma~\ref{lemma:pebe}.
\[
   \abs{\perim{\poly}} + e_P(\poly) = \abs{\border{\poly}} + e_B(\poly).
\]

Our first goal is to express the excess of a polyhex~$\poly$ as a function of the numbers of
cells of~$\poly$ of each pattern.  We denote the number of cells of a specific
pattern in~$\poly$ by $\#\hexagon$, where `$\hexagon$' is one of the~22 patterns listed
in Figure~\ref{fig:patterns_hex}.  The excess (either border or perimeter excess) of
Pattern~$\hexagon$ is denoted by $e(\hexagon)$.
(For simplicity, we omit the dependency on~$\poly$ in the notations of~$\#\hexagon$
and~$e(\hexagon)$.  This should be understood from the context.)
The border excess can be expressed
as~$e_B(\poly) = \sum_{\hexagon \in \{a,\dots,l\}} e(\hexagon)\#\hexagon$, and, similarly, the
perimeter excess can be expressed
as~$e_P(\poly) = \sum_{\hexagon \in \{o,\dots,z\}} e(\hexagon)\#\hexagon$. 
By plugging these equations into \myeqref{eq:pebe}, we obtain that
\begin{equation}
   \label{eq:all-patterns}
   \abs{\perim{\poly}} + \sum_{\hexagon \in \{o,\dots,z\}} e(\hexagon)\#\hexagon =
      \abs{\border{\poly}} + \sum_{\hexagon \in \{a,\dots,l\}} e(\hexagon)\#\hexagon~.
\end{equation}

The next step of proving the second premise is showing that minimal-perimeter polyhexes
cannot contain some of the~22 patterns.  This will simplify \myeqref{eq:all-patterns}.

\begin{lemma}
   \label{lemma:no-holes_hex}
   (Analogous to Lemma~\ref{lemma:no-holes-sqr}.)
   A minimal-perimeter polyhex does not contains holes.
\end{lemma}

\begin{proof}
   Assume to the contrary that there exists a minimal-perimeter polyhex~$\poly$ that contains
   one or more holes, and let~$\poly'$ be the polyhex obtained by filling one of the holes
   in~$\poly$.
   Clearly, $|\poly'| > |\poly|$, and by filling the hole we eliminated some perimeter cells
   and did not create new perimeter cells.
   Hence, $\abs{\perim{\poly'}} < \abs{\perim{\poly}}$.
   This contradicts the fact that~$\minp(n)$ is monotone increasing, as implied
   by Theorem~\ref{thm:minp_hex}.
   \myqed
\end{proof}

Another important observation is that minimal-perimeter polyhexes tend to be ``compact.''
We formalize this observation in the following lemma.

Recall the definition of a bridge from Section~\ref{sec:polyominoes}:
A \emph{bridge} is a cell whose removal unites two holes or renders the polyhex
disconnected (specifically, Patterns~(b), (d), (e), (g), (h), (j), and~(k)).
Similarly, a \emph{perimeter bridge} is an empty cell whose addition to the polyhex creates
a hole in it (specifically, Patterns~(p), (r), (s), (u), (v), (x),
and~(y)).

\begin{lemma}
   \label{lemma:bridges}
    (Analogous to Lemma~\ref{lemma:no-bridges-sqr}.)
   Minimal-perimeter polyhexes contain neither bridges nor perimeter bridges.
   \myqed
\end{lemma}

\begin{proof}
   Let~$\poly$ be a minimal-perimeter polyhex, and assume first that it contains a bridge cell~$f$.
   By Lemma~\ref{lemma:no-holes_hex}, since~$\poly$ does not contain holes, the removal of~$f$
   from~$\poly$ will break it into two or three disconnected polyhexes.
   We can connect these parts by translating one of them towards the other(s) by one cell.
   (In case of Pattern~(h), the polyhex is broken into three parts, but then translating
   any of them towards the removed cell would make the polyhex connected again.)
   Locally, this will eliminate at least two perimeter cells created by the bridge.
   (This can be verified by exhaustively checking all the relevant patterns.)
   The size of the new polyhex, $\poly'$, is one less than that of~$\poly$, while the
   perimeter of~$\poly'$ is smaller by at least two than that of~$\poly$.
   However, Theorem~\ref{thm:minp_hex} implies that~$\minp(n)-\minp(n-1) \leq 1$
   for all $n \geq 3$, which is a contradiction to~$\poly$ being a minimal-perimeter polyhex.
   
   Assume now that~$\poly$ contains a perimeter bridge.  Filling the bridge will not
   increase the perimeter.  (It might create one additional perimeter cell, which will be
   canceled out with the eliminated (perimeter) bridge cell.)
   In addition, it will create a hole in the polyomino.
   Then, filling the hole will create a polyhex with a larger size and a
   smaller perimeter, which is a contradiction to~$\minp(n)$ being monotone
   increasing.
   \myqed
\end{proof}

As a consequence of Lemma~\ref{lemma:no-holes_hex}, Pattern~(o) cannot appear in any
minimal-perimeter polyhex.
In addition, Lemma~\ref{lemma:bridges} tells us that the Border Patterns~(b),
(d), (e), (g), (h), (j), and~(k), as well as the Perimeter Patterns~(p),
(r), (s), (u), (v), (x), and~(y) cannot appear in any minimal-perimeter polyhex.
(Note that patterns~(b) and~(p) are not bridges by themselves, but the adjacent cell is a bridge,
that is, the cell above the central cells in~\drawpolyhex[scale=0.3]{b1.txt}
and~\drawpolyhex[scale=0.3]{p1.txt} are bridges.)
Finally, Pattern~(a) appears only in the singleton cell (the unique polyhex of size~1),
which can be disregarded.
Ignoring all these patterns, we obtain that
\begin{equation}
   \label{eq:pb321}
   \abs{\perim{\poly}} + 3\#q + 2\#t + \#w = \abs{\border{\poly}} + 3\#c + 2\#f + \#i.
\end{equation}
Note that Patterns~(l) and~(z) have excess~0, and, hence, although they may appear in
minimal-perimeter polyhexes, they do not contribute to the equation.

Consider a polyhex which contains only the six feasible patterns that contribute to the excess
(those that appear in \myeqref{eq:pb321}).
Let~$\xi$ denote the single polygon bounding the polyhex.
We now count the number of vertices and the sum of internal angles of~$\xi$ as functions of the
numbers of appearances of the different patterns.
In order to calculate the number of vertices of~$\xi$, we first determine the
number of vertices contributed by each pattern.  In order to avoid multiple counting of a
vertex, we associate each vertex to a single pattern.  Note that each vertex of~$\xi$
is surrounded by three (either occupied or empty) cells,
out of which one is empty and two are occupied, or vise versa.  We call the cell, whose type
(empty or occupied) appears once (among the surrounding three cells), the ``representative''
cell, and count only these representatives.  Thus, each vertex is counted exactly once.

For example, out of the six vertices surrounding Pattern~(c), five vertices belong to the
bounding polygon, but the representative cell of only three of them is the cell at the center
of this pattern, thus, by our scheme, Pattern~(c) contributes three vertices, each having
a~$2\pi/3$ angle.
Similarly, only two of the four vertices in the configuration of Pattern~(t), are
represented by the cell at the center of this pattern.  In this case, each vertex is the
head of a $4\pi/3$ angle.
To conclude, the total number of vertices of~$\xi$ is 
\[
   3\#c+2\#f+\#i+3\#q+2\#t+\#w,
\]
and the sum of internal angles is
\begin{equation}
   \label{eq:sum-1}
   (3\#c+2\#f+\#i)2\pi/3 + (3\#q+2\#t+\#w)4\pi/3.
\end{equation}
On the other hand, it is known that the sum of internal angles is equal to
\begin{equation}
   \label{eq:sum-2}
   (3\#c+2\#f+\#i+3\#q+2\#t+\#w-2)\pi.
\end{equation}
Equating the terms in Formulae~\eqref{eq:sum-1} and~\eqref{eq:sum-2}, we obtain that
\begin{equation}
   \label{eq:sum-3}
    3\#c+2\#f+\#i = 3\#q+2\#t+\#w + 6.
\end{equation}
Plugging this into \myeqref{eq:pb321}, we conclude
that~$\abs{\perim{\poly}} = \abs{\border{\poly}} + 6$, as required.

We also need to show that the second part of the second premise holds, that is, that if~$\poly$
is a minimal-perimeter polyhex, then $\abs{\perim{I(\poly)}} \leq \abs{\perim{\poly}} + 6$.
To this aim, note that $\border{I(\poly)} \subset \perim{\poly}$, thus, it is sufficient
to show that~$\abs{\perim{I(\poly)}} \leq \abs{\border{I(\poly)}} + 6$.  Obviously,
\myeqref{eq:all-patterns} holds for the polyhex~$I(\poly)$, hence, in order to prove the
relation, we only need to prove the following lemma.

\begin{lemma}
   \label{lemma:inf-no-bridges}
   If~$\poly$ is a minimal-perimeter polyhex, then~$I(\poly)$ does not contain any
   bridge.
   \myqed
\end{lemma}

\begin{figure}
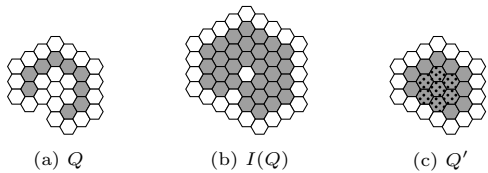

   \centering
   \begin{subfigure}[t]{0.2\textwidth}
      \centering
      \drawpolyhex[scale=0.5]{noj1.txt}
      \caption{$Q$}
   \end{subfigure}
   \begin{subfigure}[t]{0.2\textwidth}
      \centering
      \drawpolyhex[scale=0.5]{noj2.txt}
      \caption{$I(Q)$}
   \end{subfigure}
   \begin{subfigure}[t]{0.2\textwidth}
      \centering
      \drawpolyhex[scale=0.5]{noj3.txt}
      \caption{$Q'$}
   \end{subfigure}
   \caption{The construction in Lemma~\ref{lemma:inf-no-bridges} which shows that~$I(\poly)$
            cannot contain a cell of Pattern~$(j)$.  Assuming that it does, by filling the
            hole in it, we obtain~$\poly'$ which contradicts the perimeter-minimality
            of~$\poly$.  (The marked cells in~$\poly'$ are those added to~$\poly$.)}
   \label{fig:no-j}
\end{figure}

\begin{proof}
   Assume to the contrary that~$I(\poly)$ contains a bridge.
   Then, the cell that makes the bridge must have been created in the inflation
   process.  However, any cell~$c \in I(\poly) \backslash \poly$ must have a
   neighboring cell~$c' \in \poly$.  All the cells adjacent to~$c'$ must also be part
   of~$I(\poly)$, thus, cell~$c$ must have three consecutive neighbors around it,
   namely, $c'$ and the two cells neighboring both~$c$ and~$c'$.
   The only bridge pattern that fits this requirement is Pattern~(j).
   However, this means that there must have been a gap of two cells in~$\poly$ that
   caused the creation of~$c$ during the inflation of~$\poly$.  Consequently, by
   filling the gap and the hole it created, we will obtain (see Figure~\ref{fig:no-j})
   a larger polyhex with a smaller perimeter, which contradicts the fact that~$\poly$
   is a minimal-perimeter polyhex.
   \myqed
\end{proof}

\subsection{Premise 3:  Deflation Resistance}

We now show that deflating a minimal-perimeter polyhex
results in another (smaller) valid polyhex.
The intuition behind this condition is that a minimal-perimeter polyhex is ``compact,''
having a shape which does not become disconnected by deflation.

\begin{lemma}
   \label{lemma:def-valid-polyhex}
   For any minimal-perimeter polyhex~$\poly$, the shape~$D(\poly)$ is also a valid
   (connected) polyhex.
   \myqed
\end{lemma}

\begin{proof}
   The proof of this lemma is very similar to the first part of the proof of
   Lemma~\ref{lemma:bridges}.
   Consider a minimal-perimeter polyhex~$\poly$.
   In order for~$D(\poly)$ to be disconnected, $\poly$ must contain a bridge of either a
   single cell or two adjacent cells.
   A 1-cell bridge cannot be part of~$\poly$ by Lemma~\ref{lemma:bridges}.
   The polyhex~$\poly$ can neither contain a 2-cell bridge.
   \begin{figure}
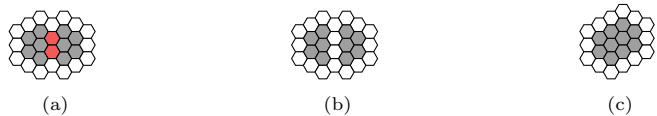

       \centering
       \begin{subfigure}[b]{0.3\textwidth}
       \centering
       \drawpolyhex[scale=0.45]{hex_bridge_1.txt}
       \caption{}
       \end{subfigure}
       \begin{subfigure}[b]{0.3\textwidth}
       \centering
       \drawpolyhex[scale=0.45]{hex_bridge_2.txt}
       \caption{}
       \end{subfigure}
       \begin{subfigure}[b]{0.3\textwidth}
       \centering
       \drawpolyhex[scale=0.45]{hex_bridge_3.txt}
       \caption{}
       \end{subfigure}
       \caption{An example for the construction in the proof of Lemma~\ref{lemma:bridges}.
                The two-cell bridge is colored in red in (a). Then, in (b), the bridge is removed,
                and, in (c), the two parts are ``glued'' together.}
       \label{fig:two-cell-bridge}
   \end{figure}
   Assume to the contrary that it does, as is shown in Figure~\ref{fig:two-cell-bridge}(a).
   Then, removing the bridge (see Figure~\ref{fig:two-cell-bridge}(b)), and then connecting
   the two pieces (by translating one of them towards the other by one cell along a direction
   which makes a $60^{\circ}$ angle with the bridge), creates (Figure~\ref{fig:two-cell-bridge}(c))
   a polyhex whose size is smaller by two than that of the original polyhex, and whose perimeter is
   smaller by at least two (since the perimeter cells adjacent to the bridge disappear).
   The new polyhex is valid, that is, the translation by one cell of one part towards the
   other does not make any cells overlap, otherwise there is a hole in the original polyhex, which
   is impossible for a minimal-perimeter polyhex by Lemma~\ref{lemma:no-holes_hex}.
   However, we reached a contradiction since for a minimal-perimeter polyhex of size~$n \geq 7$,
   we have that~$\minp(n) - \minp(n-2) \leq 1$.
   Finally, it is easy to observe by a tedious inspection that the deflation of any polyhex of
   size less than~7 results in the empty polyhex.
   \myqed
\end{proof}

\begin{figure}
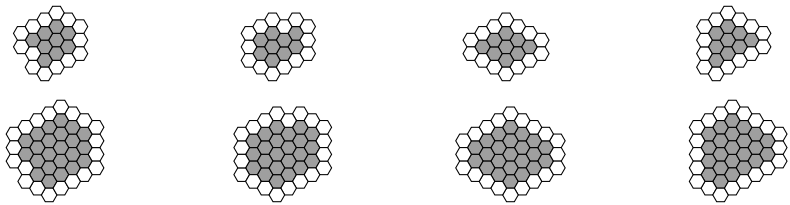

   \centering
   \begin{subfigure}[b]{0.24\textwidth}
      \centering
      \drawpolyhex[scale=0.45]{hm9_1.txt}
   \end{subfigure}
   \begin{subfigure}[b]{0.24\textwidth}
      \centering
      \drawpolyhex[scale=0.45]{hm9_2.txt}
   \end{subfigure}
   \begin{subfigure}[b]{0.24\textwidth}
      \centering
      \drawpolyhex[scale=0.45]{hm9_3.txt}
   \end{subfigure}
   \begin{subfigure}[b]{0.24\textwidth}
      \centering
      \drawpolyhex[scale=0.45]{hm9_4.txt}
   \end{subfigure} \medskip \\
   \begin{subfigure}[b]{0.24\textwidth}
      \centering
      \drawpolyhex[scale=0.45]{hm23_1.txt}
   \end{subfigure}
   \begin{subfigure}[b]{0.24\textwidth}
      \centering
      \drawpolyhex[scale=0.45]{hm23_2.txt}
   \end{subfigure} 
   \begin{subfigure}[b]{0.24\textwidth}
      \centering
      \drawpolyhex[scale=0.45]{hm23_3.txt}
   \end{subfigure}
   \begin{subfigure}[b]{0.24\textwidth}
      \centering
      \drawpolyhex[scale=0.45]{hm23_4.txt}
   \end{subfigure}
   \caption{A demonstration of Theorem~\ref{thm:main} for polyhexes.
            The top row contains all polyhexes in~$M_9$ (minimal-perimeter polyhexes of
            size~9), while the bottom row contains their
            inflated versions, all the members of~$M_{23}$.}
   \label{fig:hex_corrolary}
\end{figure}

In conclusion, we have shown that all the premises of Theorem~\ref{thm:main} are satisfied
for the hexagonal lattice, and, therefore, inflating a set of all the minimal-perimeter
polyhexes of a certain size yields another set of minimal-perimeter polyhexes of another,
larger, size.  This result is demonstrated in Figure~\ref{fig:hex_corrolary}.

\begin{figure}
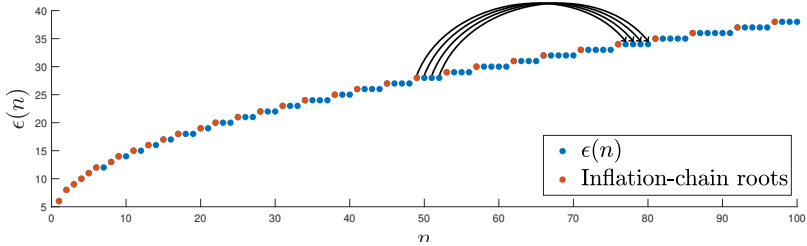

    \centering
    \includestandalone[scale=0.40]{minpH_roots}
    \vspace{-0.25cm}
    \caption{The relation between the minimum perimeter of polyhexes, $\minp(n)$, and
             the inflation-chain roots.  The points represent the minimum perimeter of a
             polyhex of size~$n$, and sizes which are inflation-chain roots are colored in red.
             The arrows show the mapping between sizes of minimal-perimeter polyhexes (induced
             by the inflation operation) and demonstrate the proof of
             Theorem~\ref{thm:root-candidates}.}
    \label{fig:minpH_roots}
\end{figure}

We also characterized inflation-chain roots of polyhexes.
As is mentioned above, the premises of Theorems~\ref{thm:main} and~\ref{thm:root-conditioned}
are satisfied for polyhexes~\cite{VainsencherB08,sieben2008polyominoes}, and, thus, the
inflation-chain roots are those who have the minimum size for a given minimal-perimeter size.
An easy consequence of Theorem~\ref{thm:minp_hex} is that the
formula~$\floor{\frac{(p-4)^2}{12}+\frac{5}{4}}$ generates all these inflation-chain roots.
This result is demonstrated in Figure~\ref{fig:minpH_roots}.

As in the case of polyominoes, and as was mentioned earlier,
it was already proven elsewhere~\cite{VainsencherB08,fulep2010polyiamonds}
that Premise~4 (roots of inflation chains) is fulfilled for the hexagonal lattice.
Therefore, we proceed to showing that Premise~5 holds.

\subsection{Premise 5:  Convergence to a Minimum-Perimeter Polyomino}

Similarly to polyominoes, we now show that starting from a polyhex~$\poly$ and applying repeatedly
a finite number, $k$, of inflation steps, we obtain a polyhex $\poly^k=I^k(\poly)$, for which
$\perim{I(\poly^k)} = \perim{\poly^k} + 6$.
Let~$R(\poly)$ denote the \emph{diameter} of~$\poly$, \emph{i.e.}, the maximal distance between two
cells of~$\poly$ when projected onto one of the three main axes.
As in the case of polyominoes,
some geometric features of~$\poly$ will disappear after $R(\poly)$ inflation steps.

\begin{lemma}
   \label{lem:no-hdz-hex}
   (Analogous to Lemma~\ref{lem:no-hdz}.)
   For any $k > R(Q)$, the polyhex~$Q^k$ does not contain any
   (i)~holes; (ii)~polyhex bridge cells; or (iii)~perimeter bridge cells.
\end{lemma}
\begin{proof}

   \begin{itemize}
   \item[(i)]
   The proof is identical to the proof for polyominoes.

   \item[(ii)]
   After~$R(Q)$ inflation steps, the obtained polyhex is clearly connected.
   If at this point there exists a bridge cell, then it must have been created in the
   last inflation step since after further steps, this cell would cease being a bridge cell.
   If during this inflation step, that eliminates the mentioned bridge,
   another bridge is created
   then its removal will not render the polyomino disconnected (since it was already connected
   before applying the inflation step), thus, it must have created a hole in the polyhex, in
   contradiction to the previous clause.

   \item[(iii)]
   We will present here a version of the analogue proof for polyominoes, adapted for polyhexes.
   Let~$c$ be a perimeter bridge cell of~$Q^k$.  Assume, without loss of generality, that two of the polyhex
   cells adjacent to it are above and below it, and denote them by~$c_1$ and~$c_2$, respectively.
   The cell whose inflation resulted in adding~$c_1$ to the
   polyhex~$c_1$,
   denoted by~$c_o$, must reside above~$c$, otherwise, it would be closer to~$c$ than to~$c_1$,
   and~$c$ would not be a perimeter cell.
   The same holds for~$c_2$ (below $c$), thus, any perimeter bridge cell must
   reside between two original cells of~$Q$.  Hence, after~$R(Q)$ inflation steps, all such cells
   will become a polyhex cells.
   \end{itemize}
\end{proof}

\begin{lemma}
   \label{lem:conv-pb4-hex}
   (Analogous to Lemma~\ref{lem:conv-pb4}.)
   After~$k = R(Q)$ inflation steps, the polyhex~$Q^k$ will obey
   $\abs{\perim{Q^k}} = \abs{\border{Q^k}} +6$.
\end{lemma}

\begin{proof}
   This follows at once from Lemma~\ref{lem:no-hdz-hex} and
   Equation~\ref{eq:sum-3}.
\end{proof}


\section{Polyiamonds}
\label{sec:polyiamonds}

\begin{figure}
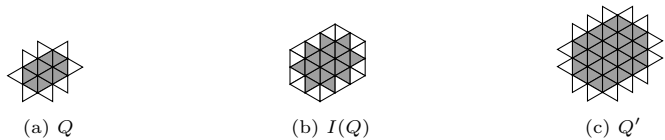

  \centering
  \begin{subfigure}[b]{0.3\textwidth}
      \centering
      \drawpolyiamond[scale=0.4]{exmp_diamond.txt}
      \caption{$\poly$}
  \end{subfigure}
  \begin{subfigure}[b]{0.3\textwidth}
      \centering
      \drawpolyiamond[scale=0.4]{exmp_diamond_I.txt}
      \caption{$I(\poly)$}
  \end{subfigure}
  \begin{subfigure}[b]{0.3\textwidth}
      \centering
      \drawpolyiamond[scale=0.4]{exmp_diamond_II.txt}
      \caption{$\poly'$}
  \end{subfigure}
  \caption{An example of inflating polyiamonds.
            The polyiamond~$\poly$ is of a minimum perimeter, however, its inflated
            version, $I(\poly)$ is not of a minimum perimeter.
            The polyiamond~$\poly'$, obtained by adding to~$\poly$ all the cells
            sharing a \emph{vertex} with $\poly$, is a minimal-perimeter polyiamond.}
  \label{fig:exmp_diamond}
\end{figure}

Polyiamonds are sets of edge-connected triangles on the regular triangular lattice.
Unlike the square and the hexagonal lattice, in which all cells are identical in shape and
in their role, the triangular lattice has two types of cells, which are seen as a left and a
right pointing arrows (\drawpolyiamond[scale=0.4]{t2_diam.txt},\drawpolyiamond[scale=0.4]{t1_diam.txt}).
Due to this complication, inflating a minimal-perimeter polyiamond does not necessarily
result in a minimal-perimeter polyiamond.  Indeed, the second premise of 
Theorem~\ref{thm:main} does not hold for polyiamonds.
This fact is not surprising, since inflating minimal-perimeter polyiamonds creates ``jaggy''
polyiamonds whose perimeter is not minimal.
Figures~\ref{fig:exmp_diamond}(a,b) illustrate this phenomenon.

However, we can fix this situation in the triangular lattice by modifying the definition
of the perimeter of a polyiamond so that it it would include all cells that share
a \emph{vertex} (instead of an edge) of the boundary of the polyiamond.
Under the new definition, Theorem~\ref{thm:main} holds.
The reason for this is surprisingly simple:
The modified definition merely mimics the inflation of animals on the graph dual to that
of the triangular lattice.  (Recall that graph duality maps vertices to faces (cells), and
vice versa, and edges to edges.)
However, the dual of the triangular lattice is the hexagonal lattice, for which we have
already shown in Section~\ref{sec:polyhexes} that all the premises of
Theorem~\ref{thm:main} hold.  Thus, applying the modified inflation operator in the
triangular lattice induces a bijection between sets of minimal-perimeter polyiamonds.
This relation is demonstrated in Figure~\ref{fig:exmp_diamond}.

\section{Conclusion}
\label{sec:conclusion}

In this paper, we show that the inflation operation induces a bijection between sets of
minimal-perimeter animals on any lattice which satisfies three conditions.
We demonstrate this result on three planar lattices:  the square, hexagonal, and also the
triangular (with a modified definition of the perimeter).
The most important contribution of this paper is the application of our result to polyhexes.
Specifically, the phenomenon of the number of isomers of a benzenoid hydrocarbons
remaining unchanged under circumscribing, which was observed in the literature of chemistry
more than~30 years ago but has never been proven till now.

However, we do not believe that this set of conditions is necessary.
Empirically, it seems that by inflating all the minimal-perimeter polycubes (animals on the
3-dimensional cubical lattice) of a given size, we obtain all the minimal-perimeter polycubes
of some larger size.  However, the second premise of Theorem~\ref{thm:main} does not hold for
this lattice.
Moreover, we believe that as stated, Theorem~\ref{thm:main} applies only to 2-dimensional
lattices! 
A simple conclusion from Lemma~\ref{lemma:pnc_size} is that if the premises of
Theorem~\ref{thm:main} hold for animals on a lattice~$\lattice$,
then~$\minp_\lattice(n) = \Theta(\sqrt{n})$.
We find it is reasonable to assume that for a $d$-dimensional lattice $\lattice_d$, the
relation between the size of a minimal-perimeter animal and its perimeter is roughly equal
to the relation between a $d$-dimensional sphere and its surface area.
Hence, we conjecture that $\minp^{\lattice_d}(n) = \Theta(n^{1-1/d})$, and, thus,
Theorem~\ref{thm:main} is not suitable for higher dimensions.



\bibliographystyle{abbrv}
\bibliography{references}

\end{document}

***************************************  LLNCS version  ***********************************

\newif \ifShowComments \ShowCommentstrue   

\documentclass{llncs}
\usepackage{microtype}
\usepackage{amsmath}
\usepackage{color}
\usepackage{cite}
\usepackage{pifont}
\usepackage{enumitem}
\usepackage{amssymb}
\usepackage{tikz}


\usepackage{subcaption}
\usepackage{polylib}

\newif\ifusestandalone
\usestandalonetrue 
\ifusestandalone
	\usepackage{standalone}
\fi


\graphicspath{{./graphics/}}
\polypath{./graphics/}

\newcommand*{\figref}[1]{\figurename~\ref{#1}}
\newcommand*{\figrefs}[1]{Figures~\ref{#1}}
\newcommand*{\myeqref}[1]{Equation~\eqref{#1}}
\newcommand{\abs}[1]{\left|#1\right|}
\newcommand{\N}{\mathbb{N}}
\newcommand{\Z}{\mathbb{Z}}
\newcommand{\hex}{\mathcal{H}}
\newcommand{\squ}{\mathcal{S}}
\newcommand{\set}[1]{\left\lbrace#1\right\rbrace}
\newcommand{\bra}[1]{\left[#1\right]}
\newcommand{\ceil}[1]{\left\lceil#1\right\rceil}
\newcommand{\floor}[1]{\left\lfloor#1\right\rfloor}

\newcommand\polylog{{\rm polylog}}

\newcommand{\perim}[1]{\mathcal{P}(#1)}
\newcommand{\border}[1]{\mathcal{B}(#1)}
\newcommand{\minp}{\epsilon}
\newcommand{\minn}{\minp^{-1}}
\newcommand{\psize}{e_P(Q)}
\newcommand{\bsize}{e_B(Q)}
\newcommand{\poly}{Q}
\newcommand{\lattice}{\mathcal{L}}
\newcommand{\hexagon}{\drawpolyhex[scale=0.6]{single_hex.txt}}

\newcommand{\myqed}{\hfill $\Box$}    
\newcommand{\comment}[1]{\relax}
\captionsetup{compatibility=false}
\captionsetup[subfigure]{justification=centering}

\newtheorem{observation}[theorem]{Observation}

\newcommand{\caplabel}[1]{#1}

\ifShowComments
\newcommand\red[1]{{\color{red}#1}}
\newcommand\blue[1]{{\color{blue}#1}}
\newcommand\purple[1]{{\color{purple}#1}}
\newcommand\toreview[1]{{\bfseries\color{olive}#1}}
\newcommand{\xnote}[3]{#1{[#2: \textbf{#3}]}}
\def\gil#1{{\xnote{\blue}{GBS says}{#1}}}
\def\gill#1{{\xnote{\red}{GB says}{#1}}}
\else
\newcommand\toreview[1]{#1}
\def\gil#1{}
\def\gill#1{}
\fi


 
\begin{document}




\frontmatter          
\pagestyle{headings}  
\addtocmark{On Minimal-Perimeter Lattice Animals}
\mainmatter              
\title{Minimal-Perimeter Lattice Animals and the Constant-Isomer Conjecture}
\date{}
%

\author{
   Gill Barequet \quad $\bullet$ \quad
   Gil Ben-Shachar
}
\authorrunning{Barequet, Ben-Shachar, Bui, and Osegueda}
\tocauthor{
   Gill Barequet (Technion, Haifa),
   Gil Ben-Shachar (Technion, Haifa)
}

\institute{
   Dept.\ of Computer Science,
   The Technion---Israel Inst.\ of Technology, \\
   Haifa~3200003, Israel.
   E-mail: \texttt{\{barequet,gilbe\}@cs.technion.ac.il}
}
\maketitle              


\begin{abstract}
   We consider minimal-perimeter lattice animals, providing a set of conditions which are sufficient
   for a lattice to have the property that inflating all minimal-perimeter animals of a
   certain size yields (without repetitions) all minimal-perimeter animals of a new, larger
   size.  We demonstrate this result on the two-dimensional square and hexagonal lattices.
   In addition, we characterize the sizes of minimal-perimeter animals on these lattices that are not
   created by inflating members of another set of minimal-perimeter animals.
\end{abstract}


\mbox{} \vspace{-12mm} \\
\begin{center}
\fbox{\includegraphics[scale=0.8]{graphics/quote.png}}
\vspace{2mm} \\
\begin{minipage}{5in}
  \footnotesize
  Cyvin S.J., Cyvin B.N., Brunvoll J. (1993) Enumeration of benzenoid chemical isomers with
  a study of constant-isomer series. In: \emph{Computer Chemistry}, part of \emph{Topics in
  Current Chemistry} book series, vol.\ 166.  Springer, Berlin, Heidelberg.  (p.~117)
\end{minipage}
\end{center}



\section{Introduction}

An \emph{animal} on a $d$-dimensional lattice is a connected set of lattice cells, where 
connectivity is through ($d{-}1$)-dimensional faces of the cells.  Specifically, on the planar
square lattice, connectivity of cells is through edges.  Two animals are considered identical if
one can be obtained from the other by \emph{translation} only, without rotations or flipping.
(Such animals are called ``fixed'' animals, as opposed to ``free'' animals.)

Lattice animals attracted interest in the literature as combinatorial objects~\cite{eden1961two} and
as a computational model in statistical physics and chemistry~\cite{temperley1956combinatorial}.
(In these areas, one usually considers \emph{site} animals, that is, clusters of lattice
vertices, hence, the graphs considered there are the \emph{dual} of our graphs.)
In this paper, we consider lattices in two dimensions, specifically, the hexagonal, triangular,
and square lattices, where animals are called polyhexes, polyiamonds, and
polyominoes, respectively.
We show the application of our results to the square and hexagonal lattices,
and explain how to extend the latter to the triangular lattice.
An example of such animals is shown in figure~\ref{fig:examples}.
\begin{figure}
    \centering
    \begin{subfigure}[t]{0.3\textwidth}
    \centering
    \drawpoly[scale = 0.75]{exmpSqr.txt}
    \end{subfigure}
    \begin{subfigure}[t]{0.3\textwidth}
    \centering
    \drawpolyhex[scale = 0.75]{exmpHex.txt}
    \end{subfigure}
    \begin{subfigure}[t]{0.3\textwidth}
    \centering
    \drawpolyiamond[scale = 0.70]{exmpTri.txt}
    \end{subfigure}
    \caption{An example of a polyomino, a polyhex, and a polyiamond.}
    \label{fig:examples}
\end{figure}
Let $A^\lattice(n)$ denote the number of lattice animals of size~$n$, that is, animals
composed of $n$ cells, on a lattice~$\lattice$.
A major research problem in the study of lattices is understanding the nature
of~$A^\lattice(n)$, either by finding a formula for it as a function of~$n$, or by evaluating
it for specific values of~$n$.
These problems are to this date still open for any nontrivial lattice.
Redelmeier~\cite{redelmeier1981counting} introduced the first algorithm for
counting all polyominoes of a given size, with no polyomino being generated more than once.
Later, Mertens~\cite{Mertens1990} showed that Redelmeier's algorithm can be
utilized for any lattice.
The first algorithm for counting lattice animals without generating all of them was
introduced by Jensen~\cite{jensen2000statistics}.  Using his method, the number of animals on
the 2-dimensional square, hexagonal, and triangular lattices were computed up to size~56,
46, and~75, respectively. 

An important measure of lattice animals is the size of their \emph{perimeter} (sometimes called
``site perimeter'').  The perimeter of a lattice animal is defined as the set of empty cells
adjacent to the animal cells.  This definition is motivated by percolation models in
statistical physics.  In such discrete models, the plane or space is made of small cells
(squares or cubes, respectively), and quanta of material or energy ``jump'' from a cell to
a neighboring cell with some probability.  Thus, the perimeter of a cluster determines
where units of material or energy can move to, and guide the statistical model of the flow.

\begin{figure}
    \centering
    \begin{subfigure}[t]{0.45\textwidth}
    \centering
    \drawpoly[scale = 0.45]{exmpQ.txt}
    \caption{Q}
    \end{subfigure}
    \begin{subfigure}[t]{0.45\textwidth}
    \centering
    \drawpoly[scale = 0.45]{exmpIQ.txt}
    \caption{I(Q)}
    \end{subfigure}
    \caption{A polyomino~$\poly$ and its inflated
            polyomino~$I(\poly)$.  Polyomino cells are colored gray, while
            perimeter cells are colored white.}
    \label{fig:exmp}
\end{figure}
Asinowski et al.~\cite{asinowski2017enumerating,asinowski2018polycubes} provided
formulae for polyominoes and polycubes with perimeter size close to the maximum possible.
On the other extreme reside animals with the \emph{minimum} possible perimeter size for their area.
The study of polyominoes of a minimal perimeter dates back to Wang and Wang~\cite{wang1977discrete},
who identified an infinite sequence of cells on the square lattice, the first~$n$ of
which (for any~$n$)
form a minimal-perimeter polyomino.  Later, Altshuler et
al.~\cite{altshuler2006}, and independently Sieben~\cite{sieben2008polyominoes}, studied
the closely-related problem of the \emph{maximum} area of a polyomino with~$p$ perimeter cells,
and provided a closed formula for the minimum possible perimeter of $n$-cell polyominoes.

Minimal-perimeter animals were also studied on other lattices.
For animals on the triangular lattice (polyiamonds), the main result is due to
F\"{u}lep and Sieben~\cite{fulep2010polyiamonds}, who characterized all the polyiamonds
with maximum area for their perimeter, and provided a formula for the minimum perimeter of a
polyiamond of size~$n$.
Similar results were given by Vainsencher and Bruckstein~\cite{VainsencherB08}
for the hexagonal lattice.
In this paper, we study an interesting property of minimal-perimeter animals, which relates to the
notion of the \emph{inflation} operation.  Simply put, inflating an animal is adding to it
all its perimeter cells (see Figure~\ref{fig:exmp}).
We provide a set of conditions (for a given lattice), which if it holds , then inflating all
minimal-perimeter animals of some size yields all minimal-perimeter animals of some larger size in a
bijective manner.

While this paper discusses some combinatorial properties of minimal-perimeter polyominoes,
another algorithmic question emerges from these properties, namely,
``how many minimal-perimeter polyominoes are there of a given size?''
This question is addressed in detail in a companion paper~\cite{barequet2020algorithms}.

The paper is organized as follows.
In Section~\ref{sec:main}, we provide some definitions and prove our main theorem.
In sections~\ref{sec:polyominoes} and~\ref{sec:polyhexes}, we show the application of
Section~\ref{sec:main} to polyominoes and polyhexes, respectivally.
Then, in Section~\ref{sec:polyiamonds} we explain how the same result also applies to the
regular triangular lattice.
We end in Section~\ref{sec:conclusion} with some concluding remarks.

\subsection{Polyhexes as Molecules}

In addition to research of minimal-perimeter animals in the literature on combinatorics,
there has been much more intensive research of minimal-perimeter polyhexes in the literature on organic
chemistry, in the context of the structure of families of molecules.
For example, significant amount of work dealt with molecules called \emph{benzenoid hydrocarbons}.
It is a known natural fact that molecules made of carbon atoms are structured as shapes on the
hexagonal lattice.  Benzenoid hydrocarbons are made of
carbon and hydrogen atoms only.  In such a molecule, the carbon atoms are arranged as a
polyhex, and the hydrogen atoms are arranged around the carbons atoms.

\begin{figure}
   \centering
   \begin{tabular}{ccc}
      \raisebox{0.39\height}{\includegraphics[scale=0.13]{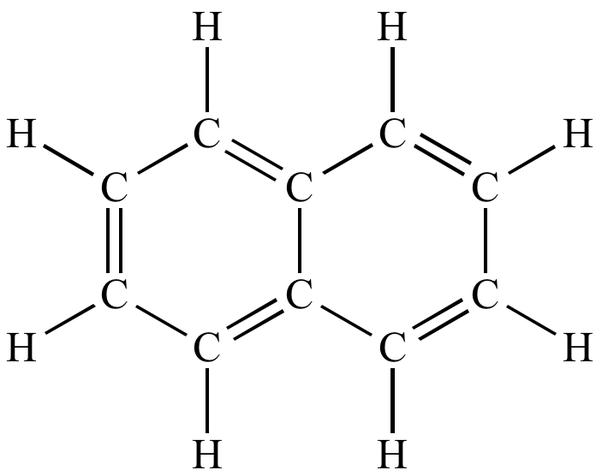}} & ~~~ &
         \includegraphics[scale=0.13]{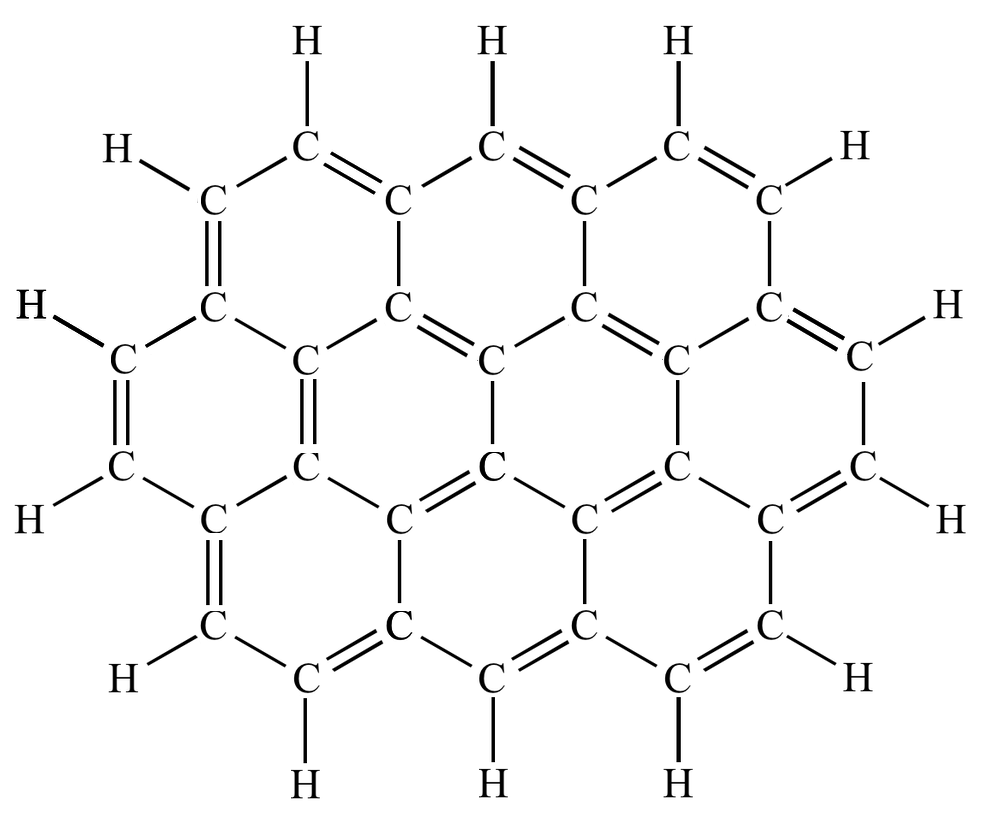} \\
      (a) Naphthalene ($C_{10} H_8$) & & (b) Circumnaphtaline ($C_{32} H_{14})$
   \end{tabular}
   \caption{Naphthalene and its circumscribed version.}
   \label{fig:naphthalene}
\end{figure}
Figure~\ref{fig:naphthalene}(a) shows a schematic drawing of the molecule of Naphthalene
(with formula~$C_{10}H_8$), the simplest benzenoid hydrocarbon, which is made of ten carbon
atoms and eight hydrogen atoms, while Figure~\ref{fig:naphthalene}(b) shows Circumnaphthalene
(molecular formula~$C_{32}H_{14}$).
There exist different configurations of atoms for the same molecular formula, which are
called \emph{isomers} of the same formula.
In the field of organic chemistry, a major goal is to enumerate all the different
isomers of a given formula.
Note that the carbon and hydrogen atoms are modeled by lattice \emph{vertices} and not by
cells of the lattice, but as we explain below, the numbers of hydrogen atoms identifies with the
number of perimeter cells of the polyhexes under discussion.
Indeed, the hydrogen atoms lie on lattice vertices that do not belong to the polyhex formed by the
carbon atoms (which also lie on lattice vertices), but are connected to them by lattice edges.
In minimal-perimeter polyhexes, each perimeter cell contains exactly two such hydrogen vertices,
and every hydrogen vertex is shared by exactly two perimeter cells.
(This has nothing to do with the fact that a single cell of the polyhex might be neighboring
several---five, in the case of Naphthalene---``empty'' cells.)
Therefore, the number of hydrogen atoms in a molecule of a benzenoid hydrocarbon is identical to the
size of the perimeter of the imaginary polyhex.\footnote{
   In order to model atoms as lattice cells, one might switch to the dual of the hexagonal lattice,
   that is, to the regular triangular lattice, but this will not serve our purpose.
}

In a series of papers (culminated in Reference~\cite{dias1987handbook}), Dias provided the
basic theory for the enumeration of benzenoid hydrocarbons.
A comprehensive review of the subject was given by Brubvoll and Cyvin~\cite{brunvoll1990we}.
Several other works~\cite{harary1976extremal,cyvin1991series,dias2010} also dealt with the
properties and enumeration of such isomers.
The analogue of what we call the ``inflation'' operation is called \emph{circumscribing} in
the literature on chemistry.
A circumscribed version of a benzenoid hydrocarbon molecule~$M$ is created by adding to~$M$
an outer layer of hexagonal ``carbon cells,'' that is, not only the hydrogen atoms (of~$M$)
adjacent to the carbon atoms now turn into carbon atoms, but also new carbon atoms are added
at all other ``free'' vertices of these cells so as to ``close'' them.
In addition, hydrogen atoms are put at all free lattice vertices that are connected by edges
to the new carbon atoms.
This process is visualized well in Figure~\ref{fig:naphthalene}.
In the literature on chemistry, it is well known that circumscribing all isomers of a given
molecular formula yields,
in a bijective manner,
all isomers that correspond to another molecular formula.
(The sequences of molecular formulae that have the same number of isomers created by
circumscribing are known as \emph{constant-isomer series}.)
Although this fact is well known, to the best of our knowledge, no rigorous proof of it
was ever given.

As mentioned above, we show that inflation induces a bijection between sets of
minimal-perimeter animals on the square, hexagonal, and in a sense, also on the triangular lattice.
By this, we prove the long-observed (but never proven)
phenomenon of ``constant-isomer series,'' that is, that circumscribing isomers of benzenoid
hydrocarbon molecules (in our terminology, inflating minimum-perimeter polyhexes)
yields all the isomers of a larger molecule.


\section{Minimal-Perimeter Animals}
\label{sec:main}

Throughout this section, we consider animals on some specific lattice~$\lattice$.
Our main result consists of a set of conditions on minimal-perimeter animals on~$\lattice$,
which is sufficient for satisfying a bijection between sets of minimal-perimeter animals on~$\lattice$.

\subsection{Preliminaries}
\label{subsec:preliminaries}

\begin{figure}
   \centering
   \begin{tabular}{ccccc}
      \drawpolyhex[scale=0.5]{exmp_hex.txt} & &
      \drawpolyhex[scale=0.5]{exmp_hex_I.txt} & &
      \drawpolyhex[scale=0.5]{exmp_hex_D.txt}\\
      $Q$ & & $I(Q)$ & & $D(Q)$
   \end{tabular}
   \caption{A polyhex~$\poly$, its inflated polyhex~$I(\poly)$, and its
            deflated polyhex~$D(\poly)$.  The gray cells belong to~$\poly$,
            the white cells are its perimeter, and
            its border cells are marked with a pattern of dots.}
    \label{fig:exmp_poly}
\end{figure}
Let~$\poly$ be an animal on~$\lattice$.
Recall that the \emph{perimeter} of~$\poly$, denoted by~$\perim{\poly}$, is the set of all
empty lattice cells that are neighbors of at least one cell of~$\poly$.
Similarly, the \emph{border} of~$\poly$, denoted by~$\border{\poly}$, is the set of cells
of~$\poly$ that are neighbors of at least one empty cell.
The \emph{inflated} version of~$\poly$ is defined as~$I(\poly) := \poly \cup \perim{\poly}$.
Similarly, the \emph{deflated} version of~$\poly$ is defined
as~$D(\poly) := \poly \backslash \border{\poly}$.
These operations are demonstrated in Figure~\ref{fig:exmp_poly}.

Denote by~$\minp(n)$ the minimum possible size of the perimeter of an $n$-cell
animal on~$\lattice$,
and by~$M_n$ the set of all minimal-perimeter $n$-cell animals on~$\lattice$.

\subsection{A Bijection}

\begin{theorem}
   \label{thm:main}
   Consider the following set of conditions.
   \begin{enumerate}[label=(\arabic*)]
      \item The function~$\minp(n)$ is weakly-monotone increasing.
      \item There exists some constant~$c^* = c^*(\lattice)$, for which, for any minimal-perimeter
            animal~$\poly$, we have that~$\abs{\perim{\poly}} = \abs{\border{\poly}} + c^*$
            and~$\abs{\perim{I(\poly)}} \leq \abs{\perim{\poly}}+c^*$.
      \item If~$\poly$ is a minimal-perimeter animal of size $n+\minp(n)$, then~$D(\poly)$ is a
            valid (connected) animal.
   \end{enumerate}
   If all the above conditions hold for~$\lattice$,
   then~$\abs{M_n} = \abs{M_{n+\minp(n)}}$.
   If these conditions are not satisfied for only a finite amount of sizes of animals, then
   the claim holds for all sizes greater than some lattice-dependent nominal size~$n_0$.
   \myqed
\end{theorem}

\begin{proof}  
   We begin with proving that inflation preserves perimeter minimality.

   \begin{lemma}
      \label{lemma:minimal-inflating}
      If~$\poly$ is a minimal-perimeter animal, then~$I(\poly)$ is a
      minimal-perimeter animal as well.
   \end{lemma}

   \begin{proof}
      Let~$\poly$ be a minimal-perimeter animal. Assume to the contrary
      that~$I(\poly)$ is not a minimal-perimeter animal, thus, there exists
      an animal~$\poly'$ such that $\abs{\poly'} = \abs{I(\poly)}$,
      and~$\abs{\perim{\poly'}} < \abs{\perim{I(\poly)}}$.
      By the second premise of Theorem~\ref{thm:main}, we know that
      $\abs{\perim{I(\poly)}} \leq \abs{\perim{\poly}} + c^*$, thus,
      $\abs{\perim{\poly'}} < \abs{\perim{\poly}}+c$, and
      since~$\poly'$ is a minimal-perimeter animal, we also know by the same premise
      that~$\abs{\perim{\poly'}} = \abs{\border{\poly'}}+c$, and, hence, 
      that~$\abs{\border{\poly'}} < \abs{\perim{\poly}}$.
      Consider now the animal~$D(\poly')$.
      Recall that $\abs{\poly'} = \abs{I(\poly)}=\abs{\poly}+\abs{\perim{\poly}}$, thus,
      the size of~$D(\poly')$ is at least~$\abs{\poly}+1$,
      and~$\abs{\perim{D(\poly')}} < \abs{\perim{\poly}} = \minp(n)$
      (since the perimeter of~$D(\poly')$ is a subset of the border of~$\poly'$).
      This is a contradiction to the first premise, which states that the
      sequence~$\minp(n)$ is monotone increasing.
      Hence, the animal~$\poly'$ cannot exist, and~$I(\poly)$ is a minimal-perimeter animal.
      \myqed
   \end{proof}

   We now proceed to demonstrating the effect of repeated inflation on the size of
   minimal-perimeter animals.

   \begin{lemma}
      \label{lemma:pnc_size}
      The minimum perimeter size of animals of size~$n+k\minp(n)+c^*k(k-1)/2$
      (for~$n > 1$ and any $k \in \N$) is~$\minp(n)+c^*k$.
   \end{lemma}

   \begin{proof}
      We repeatedly inflate a minimal-perimeter animal~$\poly$, whose initial size is~$n$.
      The size of the perimeter of~$\poly$ is~$\minp(n)$, thus, inflating
      it creates a new animal of size~$n+\minp(n)$, and the size of the border
      of~$I(\poly)$ is~$\minp(n)$, thus, the size of~$I(\poly)$
      is~$\minp(n) + c^*$.
      Continuing the inflation of the animal, the $k$th inflation will increase the size of the
      animal by $\minp(n) + (k-1)c^*$ and will increase the size of the perimeter by~$c^*$.
      Summing up these quantities yields the claim.
      \myqed
   \end{proof}

   Next, we prove that inflation preserves difference, that is, inflating two different
   minimal-perimeter animals (of equal or different sizes) always produces two different
   new animals.  (Note that this is not true for non-minimal-perimeter animals.)

   \begin{lemma}
      \label{lemma:different_inflating}
      Let~$\poly_1,\poly_2$ be two different minimal-perimeter animals.
      Then, regardless of whether or not~$\poly_1,\poly_2$ have the same size,
      the animals~$I(\poly_1)$ and~$I(\poly_2)$ are different as well.
   \end{lemma}

   \begin{proof}
      Assume to the contrary that $\poly = I(\poly_1) = I(\poly_2)$, that is,
      $\poly = \poly_1 \cup \perim{\poly_1} = \poly_2 \cup \perim{\poly_2}$.
      In addition, since $\poly_1 \neq \poly_2$, and since a cell cannot belong
      simultaneously to both an animal and to its perimeter, this means
      that~$\perim{\poly_1} \neq \perim{\poly_2}$.  The border of~$\poly$ is a
      subset of both~$\perim{\poly_1}$ and~$\perim{\poly_2}$, that is,
      $\border{\poly} \subset \perim{\poly_1} \cap \perim{\poly_2}$.
      Since~$\perim{\poly_1} \neq \perim{\poly_2}$, we obtain that
      either~$\abs{\border{\poly}} < \abs{\perim{\poly_1}}$
      or~$\abs{\border{\poly}} < \abs{\perim{\poly_2}}$;
      assume without loss of generality the former case.
      Now consider the animal~$D(\poly)$.
      Its size is~$\abs{\poly}-\abs{\border{\poly}}$.
      The size of~$\poly$ is~$\abs{\poly_1}+\abs{\perim{\poly_1}}$, thus,
      $\abs{D(\poly)} > \abs{\poly_1}$, and since the perimeter of~$D(\poly)$
      is a subset of the border of~$\poly$, we conclude
      that~$\abs{\perim{D(\poly)}} < \abs{\perim{\poly_1}}$.
      However, $\poly_1$ is a minimal-perimeter animal, which is a
      contradiction to the first premise of the theorem, which states
      that~$\minp(n)$ is monotone increasing.
      \myqed
   \end{proof}

   To complete the cycle, we also prove that for any minimal-perimeter
   animal~$\poly \in M_{n+\minp(n)}$, there is a minimal-perimeter source
   in~$M_n$, that is, an animal~$\poly'$ whose inflation yields~$\poly$.
   Specifically, this animal is $D(\poly)$.

   \begin{lemma}
      \label{lemma:deflating}
      For any~$\poly \in M_{n+\minp(n)}$, we also have
      that~$I(D(\poly)) = \poly$.
   \end{lemma}

   \begin{proof}
      Since~$\poly \in M_{n+\minp(n)}$, we have by
      Lemma~\ref{lemma:pnc_size} that~$\abs{\perim{\poly}} = \minp(n)+c^*$.
      Combining this with the equality~$\abs{\perim{\poly}} = \abs{\border{\poly}}+c^*$, we
      obtain that~$\abs{\border{\poly}} = \minp(n)$, thus, $\abs{D(\poly)} = n$
      and $\abs{\perim{D(\poly)}} \geq \minp(n)$.
      Since the perimeter of~$D(\poly)$ is a subset of the border of~$\poly$,
      and~$\abs{\border{\poly}} = \minp(n)$, we conclude that the perimeter
      of~$D(\poly)$ and the border of~$\poly$ are the same set of cells,
      and, hence, $I(D(\poly)) = \poly$.
      \myqed
   \end{proof}

   Let us now wrap up the proof of the main theorem.
   In Lemma~\ref{lemma:minimal-inflating} we have shown that for any
   minimal-perimeter animal~$\poly \in M_n$, we have that~$I(\poly) \in M_{n+\minp(n)}$.
   In addition, Lemma~\ref{lemma:different_inflating} states that the inflation of two different
   minimal-perimeter animals results in two other different minimal-perimeter animals.
   Combining the two lemmata, we obtain that~$\abs{M_n} \leq \abs{M_{n+\minp(n)}}$.
   On the other hand, in Lemma~\ref{lemma:deflating} we have shown that
   if~$\poly \in M_{n+\minp(n)}$, then~$I(D(\poly)) = \poly$, and, thus, for any animal
   in~$M_{n+\minp(n)}$, there is a unique source in~$M_n$ (specifically, $D(\poly)$),
   whose inflation yields~$\poly$.  Hence, $\abs{M_n} \geq \abs{M_{n+\minp(n)}}$.
   Combining the two relations, we conclude that~$\abs{M_n} = \abs{M_{n+\minp(n)}}$.
   \myqed
\end{proof} 

\subsection{Inflation Chains}

Theorem~\ref{thm:main} implies that there exist infinite chains of sets of minimal-perimeter
animals, each set obtained by inflating all members of the previous set, while the
cardinalities of all sets in a chain are equal.  Obviously, there are sets of
minimal-perimeter animals that are not created by the inflation of any other sets.
We call the size of animals in such sets an \emph{inflation-chain root}.
Using the definitions and proofs in the previous section, we are able to characterize which
sizes can be inflation-chain roots.
Then, using one more condition, which holds in the lattices
we consider, we determine which values are the actual inflation-chain roots.
To this aim, we define the pseudo-inverse function
\[
   \minp^{-1}(p) = \min\set{n \in \N \mid \minp(n) = p}.
\]
Since~$\minp(n)$ is a monotone-increasing discrete function, it is a step function,
and the value of~$\minp^{-1}(p)$ is the first point in each step.

\begin{theorem}
   \label{thm:root-candidates}
   Let~$\lattice$ be a lattice satisfying the premises of Theorem~\ref{thm:main}.
   Then, all inflation-chain roots are either~$\minp^{-1}(p)$
   or~$\minp^{-1}(p)-1$, for some $p \in \N$.
\end{theorem}

\begin{proof}
   Recall that~$\minp(n)$ is a step function, where each step represents all
   animal sizes for which the minimal perimeter is~$p$.
   Let us denote the start and end of the step representing the perimeter~$p$ by~$n_b^p$
   and~$n_e^p$, respectively.  Formally, $n_b^p = \minp^{-1}(p)$
   and~$n_e^p = \minp^{-1}(p+1)-1$.

   For each size~$n$ of animals in the step~$\bra{n_b^p,n_e^p}$, inflating a minimal-perimeter
   animal of size~$n$ results in an animal of size~$n{+}p$, and
   by~Lemma~\ref{lemma:pnc_size}, the perimeter of the inflated animal is~$p{+}c^*$.
   Thus, the inflation of animals of all sizes in the step of perimeter~$p$ yields animals
   that appear in the step of perimeter~$p{+}c^*$.
   In addition, they appear in a
   \emph{consecutive} portion of the step, specifically, the range~$\bra{n_b^p+p,n_e^p+p}$.
   Similarly, the step~$\bra{n_b^{p+1},n_e^{p+1}}$ is mapped by inflation to the
   range~$\bra{n_b^{p+1}+p+1,n_e^{p+1}+p+1}$, which is a portion of the step of~$p{+}1$.
   Note that the former range ends at~$n_e^p+p = n_b^{p+1}+p-1$, while the latter range
   starts at~$n_b^{p+1}+p+1$, thus, there is exactly one size of animals,
   specifically,~$n_b^{p+1}+p$, which is not covered by inflating animals in the
   ranges~$\bra{n_b^p+p,n_e^p+p}$ and~$\bra{n_b^{p+1},n_e^{p+1}}$.
   These two ranges represent two different perimeter sizes.  Hence, the size~$n_b^{p+1}+p$
   must be either the end of the first step, $n_e^{p+c^*}$,
   or the beginning of the second step,
   $n_b^{p+c^*+1}$. This concludes the proof.
   \myqed
\end{proof}

The arguments of the proof of Theorem~\ref{thm:root-candidates} are visualized in
Figure~\ref{fig:minpH_roots} for the case of polyhexes.
In fact, as we show below (see Theorem~\ref{thm:root-conditioned}), only the second
option exists, but in order to prove this, we also need a maximality-conservation
property of the inflation operation.

Here is another perspective for the above result.
Note that minimal-perimeter animals, with size corresponding to $n_e^{p}$ (for
some~$p \in \N$), are the largest animals with perimeter~$p$.
Intuitively, animals with the largest size, for a certain perimeter size, tend to be
``spherical'' (``round'' in two dimensions), and inflating them makes them even more spherical.
Therefore, one might expect that for a general lattice, the inflation operation will preserve
the property of animals being the largest for a given perimeter.  In fact, this has been proven
rigorously for the square lattice~\cite{altshuler2006,sieben2008polyominoes} and for the
hexagonal lattice~\cite{VainsencherB08,fulep2010polyiamonds}.
However, this also means that inflating a minimal-perimeter animal of size~$n_e^p$
yields a minimal-perimeter animal of size~$n_e^{p+c^*}$, and, thus, $n_e^p$ cannot be an
inflation-chain root.  We summarize this discussion in the following theorem.

\begin{theorem}
   \label{thm:root-conditioned}
   Let~$\lattice$ be a lattice for which the three premises of Theorem~\ref{thm:main} are
   satisfied, and, in addition, the following condition holds.
   \begin{enumerate}[label=(\arabic*)]
       \setcounter{enumi}{3}
       \item The inflation operation preserves the property of having a maximum size for a
             given perimeter.
   \end{enumerate}
   Then, the inflation-chain roots are precisely~$(\minp_\lattice)^{-1}(p)$, for all $p \in \N$.
   \myqed
\end{theorem}

\subsection{Convergence of Inflation Chains}

We now discuss the structure of inflated animals, and show that 
under a certain condition, inflating repeatedly \emph{any} animal (or
actually, any set, possibly disconnected, of lattice cells) ends up in a 
minimal-perimeter animal after a finite number of inflation steps.

Let~$I^k(Q)$ ($k>0$) denote the result of applying repeatedly~$k$ times the inflating
operator~$I(\cdot)$, starting from the animal~$Q$.  Equivalently, 
\[
   I^k(Q) = Q \cup \set{c \mid \mbox{Dist}(c,Q) \leq k},
\]
where~$\mbox{Dist}(c,Q)$ is the Lattice distance from a cell~$c$ to the animal~$Q$.
For brevity, we will use the notation $Q^k = I^k(Q)$.

Let us define the function 
\(
   \phi(Q) = \minn(\abs{\perim{Q}}) - \abs{Q}
\)
and explain its meaning.
When~$\phi(Q) \geq 0$, it counts the cells that should be added to~$Q$, with no change
to its perimeter, in order to make it a minimal-perimeter animal.
In particular, if~$\phi(Q) = 0$, then~$Q$ is a minimal-perimeter animal.
Otherwise, if~$\phi(Q) < 0$, then~$Q$ is also a minimal-perimeter animal, and $\abs{\phi(Q)}$ cells can be
removed from~$Q$ while still keeping the result a minimal-perimeter animal and without changing its perimeter.

\begin{lemma}
   \label{lemma:jumps-p-1}
   For any value of~$p$, we have that~$\minn(p+c^*)-\minn(p) = p-1$.
\end{lemma}

\begin{proof}
   Let~$Q$ be a minimal-perimeter animal with area~$n_b^p = \minn(p)$.
   The area of~$I(Q)$ is~$n_b^p+p$, thus, by Theorem~\ref{thm:main},
   $\perim{I(Q)} = p+c^*$.  The area~$n_b^{p+c^*}$ is an
   inflation-chain root, hence, the area of~$I(Q)$ cannot be~$n_b^{p+c^*}$.
   Except~$n_b^{p+c^*}$, animals of all other areas in the
   range~$[n_b^{p+c^*},\dots,n_e^{p+c^*}]$ are created by inflating
   minimal-perimeter animals with perimeter~$p$.
   The animal~$Q$ is of area~$n_b^p$, \emph{i.e.}, the area of~$I(Q)$ must
   be the minimal area from $\bra{n_b^{p+c^*},n_e^{p+c^*}}$ which is not
   an inflation-chain root.  Hence, the area of~$I(Q)$ is~$n_b^{p+c^*}+1$.
   We now equate the two expressions for the area of $I(Q)$:
   $n_b^p+p = n_b^{p+c^*}+1$.  That is, $n_b^{p+c^*}-n_b^{p} = p-1$.
   The claim follows.
\end{proof}

Using Lemma~\ref{lemma:jumps-p-1}, we can deduce the following result.

\begin{lemma}
   \label{lem:conv-step}
   If~$\abs{\perim{I(Q)}} = \abs{\perim{Q}} +c^*$,
   then~$\phi(I(Q)) = \phi(Q)-1$.
\end{lemma}

\begin{proof}
   \begin{align*}
      \phi(I(Q)) &= \minn(\abs{\perim{I(Q)}}) - \abs{I(Q)} \\
         &= \minn(\abs{\perim{Q}}+c^*) - (\abs{Q} + \abs{\perim{Q}}) \\
         &= \minn(\abs{\perim{Q}}) + \abs{\perim{Q}} -1 - \abs{Q}
               - \abs{\perim{Q}} \\
         &= \minn(\abs{\perim{Q}}) -\abs{Q} - 1 \\
         &= \phi(Q) - 1.
   \end{align*}
\end{proof}

Lemma~\ref{lem:conv-step} tells us that inflating an animal, $Q$, which satisfies
$\abs{\perim{I(Q)}} = \abs{\perim{Q}} +c^*$, reduces $\phi(Q)$ by $1$.
In other words, $I(Q)$ is ``closer'' than~$Q$ to being a minimal-perimeter animal.
This result is stated more formally in the following theorem.

\begin{theorem}
   \label{thm:convergence}
   Let~$\lattice$ be a lattice for which the four premises of Theorems~\ref{thm:main}
   and~\ref{thm:root-conditioned} are satisfied, and, in addition, the following condition holds.
   \begin{enumerate}[label=(\arabic*)]
      \setcounter{enumi}{4}
      \item For every animal~$Q$, there exists some finite number~$k_0 = k_0(Q)$, such that for
            every $k>k_0$, we have that~$\abs{\perim{Q^{k+1}}} = \abs{\perim{Q^{k}}} + c$.
   \end{enumerate}
   Then, after a finite number of inflation steps, 
   any animal becomes a minimal-perimeter animal.
\end{theorem}

\begin{proof}
   The claim follows from Lemma~\ref{lem:conv-step}.
   After~$k_0$ inflation operations, the premise of this lemma holds.
   Then, any additional inflation step will reduce~$\phi(Q)$ by~$1$ until~$\phi(Q)$ is nullified, which is
   precisely when the animal becomes a minimal-perimeter animal.
   (Any additional inflation steps would add superfluous cells, in the sense that they can be removed while keeping
   the animal a minimal-perimeter animal.)
\end{proof}


\section{Polyominoes}
\label{sec:polyominoes}

Throughout this section, we consider the two-dimensional square lattice~$\squ$, and
show that the premises of Theorem~\ref{thm:main} hold for this lattice.
The lattice-specific notation ($M_n$, $\minp(n)$, and~$c^*$) in this section refer to~$\squ$.

\subsection{Premise 1:  Monotonicity}

The function~$\minp^\squ(n)$, that gives the minimum possible size of the perimeter of a
polyomino of size~$n$, is known to be weakly-monotone increasing.
This fact was proved independently by Altshuler et al.~\cite{altshuler2006} and by
Sieben~\cite{sieben2008polyominoes}.
The latter reference also provides the following explicit formula.

\begin{theorem}
   \label{thm:minp_sqr}
   \textup{\cite[Thm.~$5.3$]{sieben2008polyominoes}}
   $\minp^\squ(n) = \ceil{\sqrt{8n-4} \,}+2$.
   \myqed 
\end{theorem}

\subsection{Premise 2:  Constant Inflation}

The second premise is apparently the hardest to show.
We will prove that it holds for~$\squ$ by analyzing the patterns which may appear on the
border of minimal-perimeter polyominoes.

Asinowski et al.~\cite{asinowski2017enumerating} defined the \emph{excess} of a perimeter cell
as the number of adjacent occupied cell minus one, and the total \emph{perimeter excess} of an
animal~$\poly$, $e_P(\poly)$, as the sum of excesses over all perimeter cells of~$\poly$.
We extend this definition to border cells, and, in a similar manner, define the \emph{excess} of
a border cell as the number of adjacent empty cells minus one, and the \emph{border excess}
of~$\poly$, $e_B(\poly)$, as the sum of excesses over all border cells of~$\poly$.

First, we establish a connection between the size of the perimeter of a polyomino to the size
of its border.
The following formula is universal for all lattice animals.

\begin{lemma}
   \label{lemma:pebe}
   For every animal~$\poly$, we have that
   \begin{equation}
      \label{eq:pebe}
      \abs{\perim{\poly}} + e_P(\poly) = \abs{\border{\poly}} + e_B(\poly).
   \end{equation}
\end{lemma}

\begin{proof}
   Consider the (one or more) rectilinear polygons bounding the animal~$\poly$.
   The two sides of the equation are equal to the total length of the polygon(s) in terms of
   lattice edges.
   Indeed, this length can be computed by iterating over either the border or the perimeter
   cells of~$\poly$.  In both cases, each cell contributes one edge plus its excess to the
   total length.  The claim follows.
   \myqed
\end{proof}

\begin{figure}
   \centering
   \begin{subfigure}[t]{0.1\textwidth}
      \centering
      \drawpoly[scale=0.5]{bs1.txt}
      \caption{}
   \end{subfigure}
   \begin{subfigure}[t]{0.1\textwidth}
      \centering
      \drawpoly[scale=0.5]{bs2.txt}
      \caption{}
   \end{subfigure}
   \begin{subfigure}[t]{0.1\textwidth}
      \centering
      \drawpoly[scale=0.5]{bs3.txt}
      \caption{}
   \end{subfigure}
   \begin{subfigure}[t]{0.1\textwidth}
      \centering
      \drawpoly[scale=0.5]{bs4.txt}
      \caption{}
   \end{subfigure}
   \qquad
   \begin{subfigure}[t]{0.1\textwidth}
      \centering
      \drawpoly[scale=0.5]{ps1.txt}
      \addtocounter{subfigure}{18} 
      \caption{}
   \end{subfigure}
   \begin{subfigure}[t]{0.1\textwidth}
      \centering
      \drawpoly[scale=0.5]{ps2.txt}
      \caption{}
   \end{subfigure}
   \begin{subfigure}[t]{0.1\textwidth}
      \centering
      \drawpoly[scale=0.5]{ps3.txt}
      \caption{}
   \end{subfigure}
   \begin{subfigure}[t]{0.1\textwidth}
      \centering
      \drawpoly[scale=0.5]{ps4.txt}
      \caption{}
   \end{subfigure}
   \caption{All possible patterns of cells, up to symmetries, with positive excess.
            The gray cells are polyomino cells, while the white cells are perimeter cells.
            The centers of the ``crosses'' are the subject cells, and the patterns show
            the immediate neighbors of these cells.
            Patterns~(a--d) exhibit excess border cells, while Patterns~(w--z) exhibit
            excess perimeter cells.
            }
   \label{fig:patterns_sqr}
\end{figure}

\begin{figure}
    \centering
   \drawpoly{patterns.txt}
    \caption{A~sample polyomino with marked patterns.}
    \label{fig:patterns-exmp}
\end{figure}

Let~$\#\square$ be the number of excess cells of a certain type in a polyomino,
where~`$\square$' is one of the symbols~$a$--$d$ and~$w$--$z$,
as classified in Figure~\ref{fig:patterns_sqr}.
Figure~\ref{fig:patterns-exmp} depicts a polyomino which includes cells of all these types.
Counting~$e_P(Q)$ and~$e_B(Q)$ as functions of the different patterns of
excess cells, we see that 
\(
   e_B(Q) = \#a + 2\#b + 3\#c + \#d
\)
and
\(
   e_P(Q) = \#w + 2\#x + 3\#y + \#z.
\)
Substituting~$e_B$ and~$e_P$ in \myeqref{eq:pebe}, we obtain that
\[
   \psize = \bsize +  \#a + 2\#b+3\#c + \#d - \#w - 2\#x-3\#y - \#z.
\]
Since Pattern~(c) is a singleton cell, we can ignore it in the general
formula. Thus, we have that
\[
   \psize = \bsize +  \#a + 2\#b + \#d - \#w - 2\#x-3\#y - \#z.
\]

We now simplify the equation above, first by eliminating the hole pattern, namely, Pattern~(y).

\begin{lemma}
   \label{lemma:no-holes-sqr}
   Any minimal-perimeter polyomino is simply connected (that is, it
   does not contain holes).
\end{lemma}

\begin{proof}
   The sequence~$\minp(n)$ is weakly-monotone increasing.\footnote{
      In the sequel, we simply say ``monotone increasing.''
   }
   Assume that there exists a minimal-perimeter polyomino~$\poly$ with a
   hole. Consider the polyomino~$\poly'$ that is obtained by filling this
   hole. The area of~$\poly'$ is clearly larger than that of~$\poly$,
   however, the perimeter size of~$\poly'$ is smaller than that of~$\poly$ since we eliminated
   the perimeter cells inside the hole but did not introduce new perimeter cells.
   This is a contradiction to~$\minp(n)$ being monotone increasing.
   \myqed
\end{proof}

Next, we continue to eliminate terms from the equation by showing some invariant related to the
turns of the boundary of a minimal-perimeter polyomino.

\begin{lemma}
   \label{lemma:sum_of_turns}
   For a simply connected polyomino, we have that
   \(
      \#a +2\#b -\#w -2\#x = 4.
   \)
\end{lemma}

\begin{proof}
   The boundary of a polyomino without holes is a simple polygon, thus,
   the sum of its internal angles is $(v-2)\pi$,
   where~$v$ is the complexity (number of vertices) of the polygon.
   Note that Pattern~(a) (resp.,~(b)) adds one (resp., two)
   $\pi/2$-vertex to the polygon.
   Similarly, Pattern~(w) (resp.~(x)) adds one (resp., two) $3\pi/2$-vertex.
   All other patterns do not involve vertices.
   Let~$L = \#a+2\#b$ and~$R =  \#w+2\#x$.
   Then, the sum of angles of the boundary polygon implies that
   $L \cdot \pi/2 + R \cdot 3\pi/2 = (L+R-2) \cdot \pi$,
   that is, $L-R = 4$. The claim follows.
   \myqed
\end{proof}
 
Finally, we show that Patterns~(d) and~(z) cannot exist in a minimal-perimeter polyomino.

We define a \emph{bridge} as a cell whose removal renders the polyomino disconnected.
Similarly, a perimeter bridge is a perimeter cell that neighbors two or more
connected components of the complement of the polyomino.
Observe that minimal-perimeter polyominoes do not contain any bridges, \emph{i.e.},
cells of Patterns~(d) or~(z).  This is stated in the following lemma.

\begin{lemma}
   \label{lemma:no-bridges-sqr}
   A minimal-perimeter polyomino does not contain any bridge cells.
\end{lemma}

\begin{proof}
     Let~$\poly$ be a minimal-perimeter polyomino. For the sake of
   contradiction, assume first that there is a cell~$f \in \perim{\poly}$
   as part of Pattern~(z). Assume without loss of generality
   that the two adjacent polyomino cells are to the left and to
   the right of~$f$. These two cells must be connected, thus, the area
   below (or above)~$f$ must form a cavity in the polyomino shape.
   Let, then, $\poly'$ obtained by adding~$f$ to~$\poly$ and filling the cavity.
   \figrefs{fig:no_z+d}(a,b) illustrate this situation.
   The cell directly above~$f$ becomes a perimeter cell, the cell~$f$
   ceases to be a perimeter cell, and at least one perimeter cell in the
   area filled below~$f$ is eliminated,
   thus,~$\abs{\perim{\poly'}} < \abs{\perim{\poly}}$
   and~$\abs{\poly'} > \abs{\poly}$,
   which is a contradiction to the sequence~$\minp(n)$ being
   monotone increasing.
   Therefore, polyomino~$\poly$ does not contain perimeter cells that
   fit Pattern~(z).
   
   \begin{figure}
      \centering
      \begin{subfigure}[t]{0.2\textwidth}
         \centering
         \drawpoly[scale=0.5, every node/.style={scale=0.6}]{no_z1.txt}
         \caption{$\poly$}
      \end{subfigure}
      \begin{subfigure}[t]{0.2\textwidth}
         \centering
         \drawpoly[scale=0.5, every node/.style={scale=0.6}]{no_z2.txt}
         \caption{$\poly'$}
      \end{subfigure}
      \qquad
      \begin{subfigure}[t]{0.2\textwidth}
         \centering
         \drawpoly[scale=0.5, every node/.style={scale=0.6}]{no_d1.txt}
         \caption{$\poly$}
      \end{subfigure}
      \begin{subfigure}[t]{0.2\textwidth}
         \centering
         \drawpoly[scale=0.5]{no_d2.txt}
         \caption{$\poly'$}
      \end{subfigure}
      \caption{Forbidden patterns for the proof of Theorem~\ref{theorem:pb4}.}
      \label{fig:no_z+d}
   \end{figure}

   Now assume for contradiction that~$\poly$ contains a cell~$f$ that forms
   Pattern~(d).  Let~$\poly'$ be the polyomino obtained from~$\poly$ by
   removing~$f$ (this will break~$\poly'$ into two separate pieces) and then
   shifting to the left the piece on the right (this will unite the two pieces
   into a new polyomino).
   \figrefs{fig:no_z+d}(c,d) demonstrate this situation.
   This operation is always valid since~$\poly$ is of minimal perimeter,
   hence, by Lemma~\ref{lemma:no-holes-sqr}, it is simply connected, and thus,
   removing~$f$ breaks~$\poly$ into two separate polyominoes with a gap of one
   cell in between.  Shifting to the left the piece on the right will not create a
   collision since this would mean that the two pieces were touching, which is not the case.
   On the other hand, the shift will eliminate the gap that was created by the
   removal of~$f$, hence, the two pieces will now form a new connected polyomino.
   The area of~$\poly'$
   is one less than the area of~$\poly$, and the perimeter of~$\poly'$ is
   smaller by at least two than the perimeter of~$\poly$, since the
   perimeter cells below and above~$f$ cease to be part of the perimeter,
   and connecting the two parts does not create new perimeter cells.
   From the formula of~$\minp(n)$, we know that
   $\minp(n)-\minp(n-1) \leq 1$ for~$n \geq 3$.
   However, $\abs{\poly} - \abs{\poly'} = 1$
   and~$\abs{\perim{\poly}} - \abs{\perim{\poly'}} = 2$, hence,
   $\poly$ is not a minimal-perimeter polyomino, which contradicts our
   assumption.
   Therefore, there are no cells in~$\poly$ that fit Pattern~(d).
   This completes the proof. \myqed
\end{proof}

We are now ready to wrap up the proof of the constant-inflation theorem.

\begin{theorem}
   \label{theorem:pb4}
   (Stepping Theorem)
   For any minimal-perimeter polyomino~$\poly$ (except the singleton cell), we have
   that $\psize=\bsize+4.$
\end{theorem}

\begin{proof}
   Lemma~\ref{lemma:sum_of_turns} tells us that~$\psize=\bsize+4+\#d-\#z$.
   By Lemma~\ref{lemma:no-bridges-sqr}, we know that $\#d = \#z = 0$.
   The claim follows at once.
   \myqed
\end{proof}

\subsection{Premise 3:  Deflation Resistance}

\begin{lemma}
   \label{lemma:def_valid}
   Let~$\poly$ be a minimal-perimeter polyomino of area~$n+\minp(n)$
   (for $n \geq 3$). Then, $D(\poly)$ is a valid (connected) polyomino.
\end{lemma}

\begin{proof}
   Assume to the contrary that~$D(\poly)$ is not connected, so that it is
   composed of at least two connected parts.
   Assume first that~$D(\poly)$ is composed of exactly two parts,
   $\poly_1$ and~$\poly_2$.
   Define the \emph{joint perimeter} of the two parts,
   $\perim{\poly_1,\poly_2}$, to be~$\perim{\poly_1} \cup \perim{\poly_2}$.
   Since~$\poly$ is a minimal-perimeter polyomino of area $n+\minp(n)$,
   we know by Theorem~\ref{theorem:pb4}
   that its perimeter size is~$\minp(n)+4$ and its
   border size is~$\minp(n)$, respectively.
   Thus, the size of~$D(\poly)$ is exactly~$n$ regardless of whether or
   not~$D(\poly)$ is connected.
   Since deflating~$\poly$ results in~$\poly_1 \cup \poly_2$,
   the polyomino~$\poly$ must have an (either horizontal, vertical, or
   diagonal) ``bridge'' of border cells which disappears by the deflation.
   The width of the bridge is at most~2, thus,
   $\abs{\perim{\poly_1} \cap \perim{\poly_2}} \leq 2$. Hence,
   $\abs{\perim{\poly_1}} + \abs{\perim{\poly_2}} - 2 \leq
       \abs{\perim{\poly_1,\poly_2}}$. 
   Since~$\perim{\poly_1,\poly_2}$ is a subset of~$\border{\poly}$,
   we have that $\abs{\perim{\poly_1,\poly_2}} \leq \minp(n)$. Therefore,
   \begin{equation}
      \label{eq:def_valid_1}
      \minp(\abs{\poly_1}) + \minp(\abs{\poly_2}) - 2 \leq \minp(n). 
   \end{equation}

   Recall that~$\abs{\poly_1} + \abs{\poly_2} = n$.
   It is easy to observe that~$\minp(\abs{\poly_1})+\minp(\abs{\poly_2})$
   is minimized when~$\abs{\poly_1}=1$ and $\abs{\poly_2} = n-1$ (or vice
   versa).  Had the function~$\minp(n)$ (shown in \figref{fig:minp_plot})
   \begin{figure}
      \centering
      \includegraphics[scale=0.4]{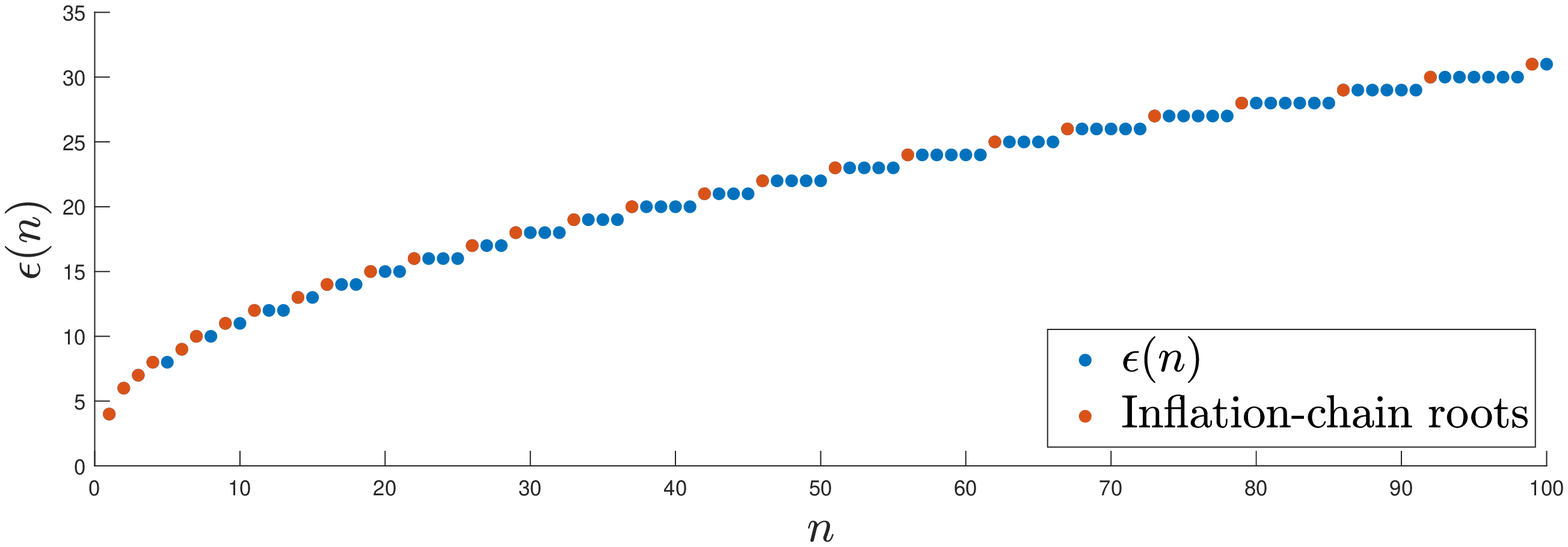}
      \caption{The function~$\minp(n)$.}
      \label{fig:minp_plot}
   \end{figure}
   been $2+\sqrt{8n-4}$ (without rounding up), this would be obvious.
   But since $\minp(n) = \left\lceil 2+\sqrt{8n-4} \, \right\rceil$, it is a
   step function (with an infinite number of intervals), where the gap
   between all successive steps is exactly~1, except the gap between the
   two leftmost steps which is~2.  This guarantees that despite the
   rounding, the minimum of~$\minp(\abs{\poly_1})+\minp(\abs{\poly_2})$
   occurs as claimed.
   Substituting this into \myeqref{eq:def_valid_1}, and using the
   fact that~$\minp(1)=4$, we see that $\minp(n-1) + 2 \leq \minp(n)$.
   However, we know~\cite{sieben2008polyominoes} that
   $\minp(n) - \minp(n-1) \leq 1$ for $n\geq 3$, which is a contradiction.
   Thus, the deflated version of~$\poly$ cannot split into two parts unless it splits into two
   singleton cells, which is indeed the case for a minimal-perimeter
   polyomino of size~8, specifically,
   \(
      D( \!\! \raisebox{-2.25mm}{\drawpoly[scale=0.3]{diag8.txt}} \!\! ) = \!\!
         \raisebox{-1.5mm}{\drawpoly[scale=0.3]{diag2.txt}}
   \).

   The same method can be used for showing that~$D(\poly)$ cannot be composed
   of more then two parts.  Note that this proof does not hold for
   polyominoes of area which is not of the form~$n+\minp(n)$, but it
   suffices for the use in Theorem~\ref{thm:main}.
   \myqed
\end{proof}

As mentioned earlier, it was already proven elsewhere~\cite{altshuler2006,sieben2008polyominoes}
that Premise~4 (roots of inflation chains) is fulfilled for the square lattice.
Therefore, we proceed to showing that Premise~5 holds.

\subsection{Premise 5:  Convergence to a Minimum-Perimeter Polyomino}

In this section, we show that starting from any polyomino~$P$, and applying repeatedly some finite number
of inflation steps, we obtain a polyomino $Q = Q(P)$, for which $\perim{I(Q)} = \perim{Q} + 4$.
Let~$R(Q)$ denote the \emph{diameter} of~$Q$, \emph{i.e.}, the maximal horizontal or
vertical distance ($L^\infty$) between two cells of~$Q$.
The following lemma shows that some geometric features of a polyomino
disappear after inflating it enough times.

\begin{lemma}
   \label{lem:no-hdz}
   For any~$k > R(Q)$, the polyomino~$Q^k$ does not contain any
   (i)~holes; (ii)~cells of Type~(d); or (iii)~patterns of Type~(z).
\end{lemma}
\begin{proof}

   \begin{itemize}
   \item[(i)]
   Let~$Q$ be a polyomino, and assume that~$Q^k$ contains a hole.
   Consider a cell~$c$ inside the hole, and let~$c_u$ be the cell 
   of~$Q^k$ that lies immediately above it.  (Note that since~$c_u$
   belongs to the border of~$Q^k$, it is not a cell of~$Q$.)  Any cell
   that resides (not necessarily directly) below~$c$ is closer to~$c$ than
   to~$c_u$.  Since $c_u \in Q^k$, it ($c_u$) is closer than~$c$ to~$Q$, thus,
   there must be a cell of~$Q$ (not necessarily directly) above~$c$,
   otherwise~$c_u$ would not belong to~$Q^k$.
   The same holds for cells below, to the right, and to the left
   of~$c$, thus,~$c$ resides within the axis-aligned bounding box of
   the extreme cells of~$Q$, and after~$R(Q)$ steps,~$c$ will be
   occupied, and any hole will be eliminated.

   \item[(ii)]
   Assume that there exists a polyomino~$Q$, for which the polyomino~$Q^k$ contains a
   cell of Type~(d).
   Without loss of generality, assume that the neighbors of~$c$ reside
   to its left and to its right, and denote them by $c_\ell,c_r$, respectively.   
   Denote by~$c_o$ one of the cells whose inflation created~$c_\ell$, \emph{i.e.},
   a cell which belongs to~$Q$ and is in distance of at most~$k$ from~$c_\ell$. 
   In addition, denote by $c_u,c_d$ the adjacent perimeter cells which
   lie immediately above and below~$c$, respectively.  The cell~$c_d$ is not
   occupied, thus, its distance from~$c_o$ is~$k+1$, which means
   that~$c_o$ lies in the same row as~$c_\ell$.  Assume for contradiction
   that~$c_o$ lies in a row below~$c_\ell$.  Then, the distance between~$c_o$ and~$c_d$
   is at most~$k$, hence~$c_d$ belongs to~$Q^k$.
   The same holds for~$c_u$; thus, cell~$c_o$ must lie in the same row as~$c_\ell$.
   Similar considerations show that~$c_o$ must lie to the left of~$c_\ell$,
   otherwise~$c_d$ and~$c_u$ would be occupied.
   In the same manner, one of the cells that originated~$c_r$
   must lie in the same row as~$c_r$ on its right.
   Hence, any cell of Type~(d) have cells of~$Q$ to its right and to its left, 
   and thus, it is found inside the bounding axis-aligned bounding box of~$Q$,
   which will necessarily be filled with polyomino cells after~$R(Q)$ inflation
   steps.

   \item[(iii)]
   Let~$c$ be a Type-(z) perimeter cell of~$Q^k$.  Assume, without loss
   of generality, that the polyomino cells adjacent to it are to its
   left and to its right, and denote them by~$c_\ell$ and~$c_r$, respectively.
   Let~$c_o$ denoted a cell whose repeated inflation has added~$c_\ell$ to $Q^k$.
   (Note that~$c_o$ might not be unique.)
   This cell must lie to the left of~$c$, otherwise, it will be closer to~$c$ than to~$c_\ell$,
   and~$c$ would not be a perimeter cell.
   In addition, $c_o$ must lie in the same row as~$c_\ell$, for otherwise, by the same 
   considerations as above, one of the cells above or below~$c$ will be occupied.
   The same holds for~$c_r$ (but to its right), thus, cells of Type~(z) must
   reside between two original cells of~$Q$, \emph{i.e.}, inside the bounding box
   of~$Q$, and after~$R(Q)$ inflation steps, all cells inside this box
   will become polyomino cells.
   \end{itemize}
\end{proof}

We can now conclude that inflating a polyomino~$Q$ for~$R(Q)$ times eliminates all
holes and bridges, and, thus, the polyomino~$Q^k$ will obey the equation
$\abs{\perim{Q^k}} = \abs{\border{Q^k}} + 4$.

\begin{lemma}
   \label{lem:conv-pb4}
   Let~$Q$ be a polyomino, and let~$k = R(Q)$.  We have that
   $\abs{\perim{Q^k}} = \abs{\border{Q^k}} + 4$.
\end{lemma}

\begin{proof}
   This follows at once from Lemma~\ref{lem:no-hdz} and
   Theorem~\ref{theorem:pb4}.
\end{proof}


\section{Polyhexes}
\label{sec:polyhexes}

In this section, we show that the premises of Theorem~\ref{thm:main} hold for the
two-dimensional hexagonal lattice~$\hex$.  The roadmap followed in this section is
similar to the one used in Section~\ref{sec:polyominoes}.  In this section, all the
lattice-specific notations refer to~$\hex$.

\subsection{Premise 1:  Monotonicity}

The first premise has been proven for~$\hex$ independently by Vainsencher and
Bruckstien~\cite{VainsencherB08} and by F\"{u}lep and Sieben~\cite{fulep2010polyiamonds}.
We will use the latter, stronger version which also includes a formula for $\minp(n)$.

\begin{theorem}
   \label{thm:minp_hex}
   \textup{\cite[Thm.~$5.12$]{fulep2010polyiamonds}}
   $\minp(n) = \ceil{\sqrt{12n-3}\,}+3$.
   \myqed 
\end{theorem}

Clearly, the function~$\minp(n)$ is weakly-monotone increasing.

\subsection{Premise 2:  Constant Inflation}

To show that the second premise holds, we analyze the different patterns that may
appear in the border and perimeter of minimal-perimeter polyhexes.
We can classify every border or perimeter cell by one of exactly~24 patterns,
distinguished by the number and positions of their adjacent occupied cells.
The~24 possible patterns are shown in Figure~\ref{fig:patterns_hex}.
\begin{figure}
   \centering
   \begin{subfigure}[t]{0.075\textwidth}
      \centering
      \drawpolyhex[scale=0.65]{b0.txt}
      \caption{}
      \label{fig:b0}
   \end{subfigure}
   \begin{subfigure}[t]{0.075\textwidth}
      \centering
      \drawpolyhex[scale=0.65]{b1.txt}
      \caption{}
      \label{fig:b1}
   \end{subfigure}
   \begin{subfigure}[t]{0.075\textwidth}
      \centering
      \drawpolyhex[scale=0.65]{b2.txt}
      \caption{}
      \label{fig:b2}
   \end{subfigure}
   \begin{subfigure}[t]{0.075\textwidth}
      \centering
      \drawpolyhex[scale=0.65]{b3.txt}
      \caption{}
      \label{fig:b3}
   \end{subfigure}
   \begin{subfigure}[t]{0.075\textwidth}
      \centering
      \drawpolyhex[scale=0.65]{b4.txt}
      \caption{}
      \label{fig:b4}
   \end{subfigure}
   \begin{subfigure}[t]{0.075\textwidth}
      \centering
      \drawpolyhex[scale=0.65]{b5.txt}
      \caption{}
      \label{fig:b5}
   \end{subfigure}
   \begin{subfigure}[t]{0.075\textwidth}
      \centering
      \drawpolyhex[scale=0.65]{b6.txt}
      \caption{}
      \label{fig:b6}
   \end{subfigure}
   \begin{subfigure}[t]{0.075\textwidth}
      \centering
      \drawpolyhex[scale=0.65]{b7.txt}
      \caption{}
      \label{fig:b7}
   \end{subfigure}
   \begin{subfigure}[t]{0.075\textwidth}
      \centering
      \drawpolyhex[scale=0.65]{b8.txt}
      \caption{}
      \label{fig:b8}
   \end{subfigure}
   \begin{subfigure}[t]{0.075\textwidth}
      \centering
      \drawpolyhex[scale=0.65]{b9.txt}
      \caption{}
      \label{fig:b9}
   \end{subfigure}
   \begin{subfigure}[t]{0.075\textwidth}
      \centering
      \drawpolyhex[scale=0.65]{b10.txt}
      \caption{}
      \label{fig:b10}
   \end{subfigure}
   \begin{subfigure}[t]{0.075\textwidth}
      \centering
      \drawpolyhex[scale=0.65]{b11.txt}
      \caption{}
      \label{fig:b11}
   \end{subfigure}
   \medskip \\
   \begin{subfigure}[t]{0.075\textwidth}
      \centering
      \addtocounter{subfigure}{2} 
      \drawpolyhex[scale=0.65]{p0.txt}
      \caption{}
      \label{fig:p0}
   \end{subfigure}
   \begin{subfigure}[t]{0.075\textwidth}
      \centering
      \drawpolyhex[scale=0.65]{p1.txt}
      \caption{}
      \label{fig:p1}
   \end{subfigure}
   \begin{subfigure}[t]{0.075\textwidth}
      \centering
      \drawpolyhex[scale=0.65]{p2.txt}
      \caption{}
      \label{fig:p2}
   \end{subfigure}
   \begin{subfigure}[t]{0.075\textwidth}
      \centering
      \drawpolyhex[scale=0.65]{p3.txt}
      \caption{}
      \label{fig:p3}
   \end{subfigure}
   \begin{subfigure}[t]{0.075\textwidth}
      \centering
      \drawpolyhex[scale=0.65]{p4.txt}
      \caption{}
      \label{fig:p4}
   \end{subfigure}
   \begin{subfigure}[t]{0.075\textwidth}
      \centering
      \drawpolyhex[scale=0.65]{p5.txt}
      \caption{}
      \label{fig:p5}
   \end{subfigure}
   \begin{subfigure}[t]{0.075\textwidth}
      \centering
      \drawpolyhex[scale=0.65]{p6.txt}
      \caption{}
      \label{fig:p6}
   \end{subfigure}
   \begin{subfigure}[t]{0.075\textwidth}
      \centering
      \drawpolyhex[scale=0.65]{p7.txt}
      \caption{}
      \label{fig:p7}
   \end{subfigure}
   \begin{subfigure}[t]{0.075\textwidth}
      \centering
      \drawpolyhex[scale=0.65]{p8.txt}
      \caption{}
      \label{fig:p8}
   \end{subfigure}
   \begin{subfigure}[t]{0.075\textwidth}
      \centering
      \drawpolyhex[scale=0.65]{p9.txt}
      \caption{}
      \label{fig:p9}
   \end{subfigure}
   \begin{subfigure}[t]{0.075\textwidth}
      \centering
      \drawpolyhex[scale=0.65]{p10.txt}
      \caption{}
      \label{fig:p10}
   \end{subfigure}
   \begin{subfigure}[t]{0.075\textwidth}
      \centering
      \drawpolyhex[scale=0.65]{p11.txt}
      \caption{}
      \label{fig:p11}
   \end{subfigure}
   \caption{All possible patterns (up to symmetries) of border (first row) and perimeter
            (second row) cells.
            The gray cells are polyhex cells, while the white cells are perimeter cells.
            Each subfigure shows a cell in the middle, and the possible pattern of cells
            surrounding it.}
   \label{fig:patterns_hex}
\end{figure}

Let us recall the equation subject of Lemma~\ref{lemma:pebe}.
\[
   \abs{\perim{\poly}} + e_P(\poly) = \abs{\border{\poly}} + e_B(\poly).
\]

Our first goal is to express the excess of a polyhex~$\poly$ as a function of the numbers of
cells of~$\poly$ of each pattern.  We denote the number of cells of a specific
pattern in~$\poly$ by $\#\hexagon$, where `$\hexagon$' is one of the~22 patterns listed
in Figure~\ref{fig:patterns_hex}.  The excess (either border or perimeter excess) of
Pattern~$\hexagon$ is denoted by $e(\hexagon)$.
(For simplicity, we omit the dependency on~$\poly$ in the notations of~$\#\hexagon$
and~$e(\hexagon)$.  This should be understood from the context.)
The border excess can be expressed
as~$e_B(\poly) = \sum_{\hexagon \in \{a,\dots,l\}} e(\hexagon)\#\hexagon$, and, similarly, the
perimeter excess can be expressed
as~$e_P(\poly) = \sum_{\hexagon \in \{o,\dots,z\}} e(\hexagon)\#\hexagon$. 
By plugging these equations into \myeqref{eq:pebe}, we obtain that
\begin{equation}
   \label{eq:all-patterns}
   \abs{\perim{\poly}} + \sum_{\hexagon \in \{o,\dots,z\}} e(\hexagon)\#\hexagon =
      \abs{\border{\poly}} + \sum_{\hexagon \in \{a,\dots,l\}} e(\hexagon)\#\hexagon~.
\end{equation}

The next step of proving the second premise is showing that minimal-perimeter polyhexes
cannot contain some of the~22 patterns.  This will simplify \myeqref{eq:all-patterns}.

\begin{lemma}
   \label{lemma:no-holes_hex}
   (Analogous to Lemma~\ref{lemma:no-holes-sqr}.)
   A minimal-perimeter polyhex does not contains holes.
\end{lemma}

\begin{proof}
   Assume to the contrary that there exists a minimal-perimeter polyhex~$\poly$ that contains
   one or more holes, and let~$\poly'$ be the polyhex obtained by filling one of the holes
   in~$\poly$.
   Clearly, $|\poly'| > |\poly|$, and by filling the hole we eliminated some perimeter cells
   and did not create new perimeter cells.
   Hence, $\abs{\perim{\poly'}} < \abs{\perim{\poly}}$.
   This contradicts the fact that~$\minp(n)$ is monotone increasing, as implied
   by Theorem~\ref{thm:minp_hex}.
   \myqed
\end{proof}

Another important observation is that minimal-perimeter polyhexes tend to be ``compact.''
We formalize this observation in the following lemma.

Recall the definition of a bridge from Section~\ref{sec:polyominoes}:
A \emph{bridge} is a cell whose removal unites two holes or renders the polyhex
disconnected (specifically, Patterns~(b), (d), (e), (g), (h), (j), and~(k)).
Similarly, a \emph{perimeter bridge} is an empty cell whose addition to the polyhex creates
a hole in it (specifically, Patterns~(p), (r), (s), (u), (v), (x),
and~(y)).

\begin{lemma}
   \label{lemma:bridges}
    (Analogous to Lemma~\ref{lemma:no-bridges-sqr}.)
   Minimal-perimeter polyhexes contain neither bridges nor perimeter bridges.
   \myqed
\end{lemma}

\begin{proof}
   Let~$\poly$ be a minimal-perimeter polyhex, and assume first that it contains a bridge cell~$f$.
   By Lemma~\ref{lemma:no-holes_hex}, since~$\poly$ does not contain holes, the removal of~$f$
   from~$\poly$ will break it into two or three disconnected polyhexes.
   We can connect these parts by translating one of them towards the other(s) by one cell.
   (In case of Pattern~(h), the polyhex is broken into three parts, but then translating
   any of them towards the removed cell would make the polyhex connected again.)
   Locally, this will eliminate at least two perimeter cells created by the bridge.
   (This can be verified by exhaustively checking all the relevant patterns.)
   The size of the new polyhex, $\poly'$, is one less than that of~$\poly$, while the
   perimeter of~$\poly'$ is smaller by at least two than that of~$\poly$.
   However, Theorem~\ref{thm:minp_hex} implies that~$\minp(n)-\minp(n-1) \leq 1$
   for all $n \geq 3$, which is a contradiction to~$\poly$ being a minimal-perimeter polyhex.
   
   Assume now that~$\poly$ contains a perimeter bridge.  Filling the bridge will not
   increase the perimeter.  (It might create one additional perimeter cell, which will be
   canceled out with the eliminated (perimeter) bridge cell.)
   In addition, it will create a hole in the polyomino.
   Then, filling the hole will create a polyhex with a larger size and a
   smaller perimeter, which is a contradiction to~$\minp(n)$ being monotone
   increasing.
   \myqed
\end{proof}

As a consequence of Lemma~\ref{lemma:no-holes_hex}, Pattern~(o) cannot appear in any
minimal-perimeter polyhex.
In addition, Lemma~\ref{lemma:bridges} tells us that the Border Patterns~(b),
(d), (e), (g), (h), (j), and~(k), as well as the Perimeter Patterns~(p),
(r), (s), (u), (v), (x), and~(y) cannot appear in any minimal-perimeter polyhex.
(Note that patterns~(b) and~(p) are not bridges by themselves, but the adjacent cell is a bridge,
that is, the cell above the central cells in~\drawpolyhex[scale=0.3]{b1.txt}
and~\drawpolyhex[scale=0.3]{p1.txt} are bridges.)
Finally, Pattern~(a) appears only in the singleton cell (the unique polyhex of size~1),
which can be disregarded.
Ignoring all these patterns, we obtain that
\begin{equation}
   \label{eq:pb321}
   \abs{\perim{\poly}} + 3\#q + 2\#t + \#w = \abs{\border{\poly}} + 3\#c + 2\#f + \#i.
\end{equation}
Note that Patterns~(l) and~(z) have excess~0, and, hence, although they may appear in
minimal-perimeter polyhexes, they do not contribute to the equation.

Consider a polyhex which contains only the six feasible patterns that contribute to the excess
(those that appear in \myeqref{eq:pb321}).
Let~$\xi$ denote the single polygon bounding the polyhex.
We now count the number of vertices and the sum of internal angles of~$\xi$ as functions of the
numbers of appearances of the different patterns.
In order to calculate the number of vertices of~$\xi$, we first determine the
number of vertices contributed by each pattern.  In order to avoid multiple counting of a
vertex, we associate each vertex to a single pattern.  Note that each vertex of~$\xi$
is surrounded by three (either occupied or empty) cells,
out of which one is empty and two are occupied, or vise versa.  We call the cell, whose type
(empty or occupied) appears once (among the surrounding three cells), the ``representative''
cell, and count only these representatives.  Thus, each vertex is counted exactly once.

For example, out of the six vertices surrounding Pattern~(c), five vertices belong to the
bounding polygon, but the representative cell of only three of them is the cell at the center
of this pattern, thus, by our scheme, Pattern~(c) contributes three vertices, each having
a~$2\pi/3$ angle.
Similarly, only two of the four vertices in the configuration of Pattern~(t), are
represented by the cell at the center of this pattern.  In this case, each vertex is the
head of a $4\pi/3$ angle.
To conclude, the total number of vertices of~$\xi$ is 
\[
   3\#c+2\#f+\#i+3\#q+2\#t+\#w,
\]
and the sum of internal angles is
\begin{equation}
   \label{eq:sum-1}
   (3\#c+2\#f+\#i)2\pi/3 + (3\#q+2\#t+\#w)4\pi/3.
\end{equation}
On the other hand, it is known that the sum of internal angles is equal to
\begin{equation}
   \label{eq:sum-2}
   (3\#c+2\#f+\#i+3\#q+2\#t+\#w-2)\pi.
\end{equation}
Equating the terms in Formulae~\eqref{eq:sum-1} and~\eqref{eq:sum-2}, we obtain that
\begin{equation}
   \label{eq:sum-3}
    3\#c+2\#f+\#i = 3\#q+2\#t+\#w + 6.
\end{equation}
Plugging this into \myeqref{eq:pb321}, we conclude
that~$\abs{\perim{\poly}} = \abs{\border{\poly}} + 6$, as required.

We also need to show that the second part of the second premise holds, that is, that if~$\poly$
is a minimal-perimeter polyhex, then $\abs{\perim{I(\poly)}} \leq \abs{\perim{\poly}} + 6$.
To this aim, note that $\border{I(\poly)} \subset \perim{\poly}$, thus, it is sufficient
to show that~$\abs{\perim{I(\poly)}} \leq \abs{\border{I(\poly)}} + 6$.  Obviously,
\myeqref{eq:all-patterns} holds for the polyhex~$I(\poly)$, hence, in order to prove the
relation, we only need to prove the following lemma.

\begin{lemma}
   \label{lemma:inf-no-bridges}
   If~$\poly$ is a minimal-perimeter polyhex, then~$I(\poly)$ does not contain any
   bridge.
   \myqed
\end{lemma}

\begin{proof}
   Assume to the contrary that~$I(\poly)$ contains a bridge.
   Then, the cell that makes the bridge must have been created in the inflation
   process.  However, any cell~$c \in I(\poly) \backslash \poly$ must have a
   neighboring cell~$c' \in \poly$.  All the cells adjacent to~$c'$ must also be part
   of~$I(\poly)$, thus, cell~$c$ must have three consecutive neighbors around it,
   namely, $c'$ and the two cells neighboring both~$c$ and~$c'$.
   The only bridge pattern that fits this requirement is Pattern~(j).
   However, this means that there must have been a gap of two cells in~$\poly$ that
   caused the creation of~$c$ during the inflation of~$\poly$.  Consequently, by
   filling the gap and the hole it created, we will obtain (see Figure~\ref{fig:no-j})
   a larger polyhex with a smaller perimeter, which contradicts the fact that~$\poly$
   is a minimal-perimeter polyhex.
   \myqed
\end{proof}
\begin{figure}
   \centering
   \begin{subfigure}[t]{0.2\textwidth}
      \centering
      \drawpolyhex[scale=0.5]{noj1.txt}
      \caption{$Q$}
   \end{subfigure}
   \begin{subfigure}[t]{0.2\textwidth}
      \centering
      \drawpolyhex[scale=0.5]{noj2.txt}
      \caption{$I(Q)$}
   \end{subfigure}
   \begin{subfigure}[t]{0.2\textwidth}
      \centering
      \drawpolyhex[scale=0.5]{noj3.txt}
      \caption{$Q'$}
   \end{subfigure}
   \caption{The construction in Lemma~\ref{lemma:inf-no-bridges} which shows that~$I(\poly)$
            cannot contain a cell of Pattern~$(j)$.  Assuming that it does, by filling the
            hole in it, we obtain~$\poly'$ which contradicts the perimeter-minimality
            of~$\poly$.  (The marked cells in~$\poly'$ are those added to~$\poly$.)}
   \label{fig:no-j}
\end{figure}

\subsection{Premise 3:  Deflation Resistance}

We now show that deflating a minimal-perimeter polyhex
results in another (smaller) valid polyhex.
The intuition behind this condition is that a minimal-perimeter polyhex is ``compact,''
having a shape which does not become disconnected by deflation.

\begin{lemma}
   \label{lemma:def-valid-polyhex}
   For any minimal-perimeter polyhex~$\poly$, the shape~$D(\poly)$ is also a valid
   (connected) polyhex.
   \myqed
\end{lemma}

\begin{proof}
   The proof of this lemma is very similar to the first part of the proof of
   Lemma~\ref{lemma:bridges}.
   Consider a minimal-perimeter polyhex~$\poly$.
   In order for~$D(\poly)$ to be disconnected, $\poly$ must contain a bridge of either a
   single cell or two adjacent cells.
   A 1-cell bridge cannot be part of~$\poly$ by Lemma~\ref{lemma:bridges}.
   The polyhex~$\poly$ can neither contain a 2-cell bridge.
   \begin{figure}
       \centering
       \begin{subfigure}[b]{0.3\textwidth}
       \centering
       \drawpolyhex[scale=0.45]{hex_bridge_1.txt}
       \caption{}
       \end{subfigure}
       \begin{subfigure}[b]{0.3\textwidth}
       \centering
       \drawpolyhex[scale=0.45]{hex_bridge_2.txt}
       \caption{}
       \end{subfigure}
       \begin{subfigure}[b]{0.3\textwidth}
       \centering
       \drawpolyhex[scale=0.45]{hex_bridge_3.txt}
       \caption{}
       \end{subfigure}
       \caption{An example for the construction in the proof of Lemma~\ref{lemma:bridges}.
                The two-cell bridge is colored in red in (a). Then, in (b), the bridge is removed,
                and, in (c), the two parts are ``glued'' together.}
       \label{fig:two-cell-bridge}
   \end{figure}
   Assume to the contrary that it does, as is shown in Figure~\ref{fig:two-cell-bridge}(a).
   Then, removing the bridge (see Figure~\ref{fig:two-cell-bridge}(b)), and then connecting
   the two pieces (by translating one of them towards the other by one cell along a direction
   which makes a $60^{\circ}$ angle with the bridge), creates (Figure~\ref{fig:two-cell-bridge}(c))
   a polyhex whose size is smaller by two than that of the original polyhex, and whose perimeter is
   smaller by at least two (since the perimeter cells adjacent to the bridge disappear).
   The new polyhex is valid, that is, the translation by one cell of one part towards the
   other does not make any cells overlap, otherwise there is a hole in the original polyhex, which
   is impossible for a minimal-perimeter polyhex by Lemma~\ref{lemma:no-holes_hex}.
   However, we reached a contradiction since for a minimal-perimeter polyhex of size~$n \geq 7$,
   we have that~$\minp(n) - \minp(n-2) \leq 1$.
   Finally, it is easy to observe by a tedious inspection that the deflation of any polyhex of
   size less than~7 results in the empty polyhex.
   \myqed
\end{proof}

In conclusion, we have shown that all the premises of Theorem~\ref{thm:main} are satisfied
for the hexagonal lattice, and, therefore, inflating a set of all the minimal-perimeter
polyhexes of a certain size yields another set of minimal-perimeter polyhexes of another,
larger, size.  This result is demonstrated in Figure~\ref{fig:hex_corrolary}.

\begin{figure}
   \centering
   \begin{subfigure}[b]{0.24\textwidth}
      \centering
      \drawpolyhex[scale=0.45]{hm9_1.txt}
   \end{subfigure}
   \begin{subfigure}[b]{0.24\textwidth}
      \centering
      \drawpolyhex[scale=0.45]{hm9_2.txt}
   \end{subfigure}
   \begin{subfigure}[b]{0.24\textwidth}
      \centering
      \drawpolyhex[scale=0.45]{hm9_3.txt}
   \end{subfigure}
   \begin{subfigure}[b]{0.24\textwidth}
      \centering
      \drawpolyhex[scale=0.45]{hm9_4.txt}
   \end{subfigure} \medskip \\
   \begin{subfigure}[b]{0.24\textwidth}
      \centering
      \drawpolyhex[scale=0.45]{hm23_1.txt}
   \end{subfigure}
   \begin{subfigure}[b]{0.24\textwidth}
      \centering
      \drawpolyhex[scale=0.45]{hm23_2.txt}
   \end{subfigure} 
   \begin{subfigure}[b]{0.24\textwidth}
      \centering
      \drawpolyhex[scale=0.45]{hm23_3.txt}
   \end{subfigure}
   \begin{subfigure}[b]{0.24\textwidth}
      \centering
      \drawpolyhex[scale=0.45]{hm23_4.txt}
   \end{subfigure}
   \caption{A demonstration of Theorem~\ref{thm:main} for polyhexes.
            The top row contains all polyhexes in~$M_9$ (minimal-perimeter polyhexes of
            size~9), while the bottom row contains their
            inflated versions, all the members of~$M_{23}$.}
   \label{fig:hex_corrolary}
\end{figure}

We also characterized inflation-chain roots of polyhexes.
As is mentioned above, the premises of Theorems~\ref{thm:main} and~\ref{thm:root-conditioned}
are satisfied for polyhexes~\cite{VainsencherB08,sieben2008polyominoes}, and, thus, the
inflation-chain roots are those who have the minimum size for a given minimal-perimeter size.
An easy consequence of Theorem~\ref{thm:minp_hex} is that the
formula~$\floor{\frac{(p-4)^2}{12}+\frac{5}{4}}$ generates all these inflation-chain roots.
This result is demonstrated in Figure~\ref{fig:minpH_roots}.

\begin{figure}
    \centering
    \includestandalone[scale=0.40]{minpH_roots}
    \vspace{-0.25cm}
    \caption{The relation between the minimum perimeter of polyhexes, $\minp(n)$, and
             the inflation-chain roots.  The points represent the minimum perimeter of a
             polyhex of size~$n$, and sizes which are inflation-chain roots are colored in red.
             The arrows show the mapping between sizes of minimal-perimeter polyhexes (induced
             by the inflation operation) and demonstrate the proof of
             Theorem~\ref{thm:root-candidates}.}
    \label{fig:minpH_roots}
\end{figure}

As in the case of polyominoes, and as was mentioned earlier,
it was already proven elsewhere~\cite{VainsencherB08,fulep2010polyiamonds}
that Premise~4 (roots of inflation chains) is fulfilled for the hexagonal lattice.
Therefore, we proceed to showing that Premise~5 holds.

\subsection{Premise 5:  Convergence to a Minimum-Perimeter Polyomino}

Similarly to polyominoes, we now show that starting from a polyhex~$\poly$ and applying repeatedly
a finite number, $k$, of inflation steps, we obtain a polyhex $\poly^k=I^k(\poly)$, for which
$\perim{I(\poly^k)} = \perim{\poly^k} + 6$.
Let~$R(\poly)$ denote the \emph{diameter} of~$\poly$, \emph{i.e.}, the maximal distance between two
cells of~$\poly$ when projected onto one of the three main axes.
As in the case of polyominoes,
some geometric features of~$\poly$ will disappear after $R(\poly)$ inflation steps.

\begin{lemma}
   \label{lem:no-hdz-hex}
   (Analogous to Lemma~\ref{lem:no-hdz}.)
   For any $k > R(Q)$, the polyhex~$Q^k$ does not contain any
   (i)~holes; (ii)~polyhex bridge cells; or (iii)~perimeter bridge cells.
\end{lemma}
\begin{proof}

   \begin{itemize}
   \item[(i)]
   The proof is identical to the proof for polyominoes.

   \item[(ii)]
   After~$R(Q)$ inflation steps, the obtained polyhex is clearly connected.
   If at this point there exists a bridge cell, then it must have been created in the
   last inflation step since after further steps, this cell would cease being a bridge cell.
   If during this inflation step, that eliminates the mentioned bridge,
   another bridge is created
   then its removal will not render the polyomino disconnected (since it was already connected
   before applying the inflation step), thus, it must have created a hole in the polyhex, in
   contradiction to the previous clause.

   \item[(iii)]
   We will present here a version of the analogue proof for polyominoes, adapted for polyhexes.
   Let~$c$ be a perimeter bridge cell of~$Q^k$.  Assume, without loss of generality, that two of the polyhex
   cells adjacent to it are above and below it, and denote them by~$c_1$ and~$c_2$, respectively.
   The cell whose inflation resulted in adding~$c_1$ to the
   polyhex~$c_1$,
   denoted by~$c_o$, must reside above~$c$, otherwise, it would be closer to~$c$ than to~$c_1$,
   and~$c$ would not be a perimeter cell.
   The same holds for~$c_2$ (below $c$), thus, any perimeter bridge cell must
   reside between two original cells of~$Q$.  Hence, after~$R(Q)$ inflation steps, all such cells
   will become a polyhex cells.
   \end{itemize}
\end{proof}

\begin{lemma}
   \label{lem:conv-pb4-hex}
   (Analogous to Lemma~\ref{lem:conv-pb4}.)
   After~$k = R(Q)$ inflation steps, the polyhex~$Q^k$ will obey
   $\abs{\perim{Q^k}} = \abs{\border{Q^k}} +6$.
\end{lemma}

\begin{proof}
   This follows at once from Lemma~\ref{lem:no-hdz-hex} and
   Equation~\ref{eq:sum-3}.
\end{proof}


\section{Polyiamonds}
\label{sec:polyiamonds}

Polyiamonds are sets of edge-connected triangles on the regular triangular lattice.
Unlike the square and the hexagonal lattice, in which all cells are identical in shape and
in their role, the triangular lattice has two types of cells, which are seen as a left and a
right pointing arrows (\drawpolyiamond[scale=0.4]{t2_diam.txt},\drawpolyiamond[scale=0.4]{t1_diam.txt}).
Due to this complication, inflating a minimal-perimeter polyiamond does not necessarily
result in a minimal-perimeter polyiamond.  Indeed, the second premise of 
Theorem~\ref{thm:main} does not hold for polyiamonds.
This fact is not surprising, since inflating minimal-perimeter polyiamonds creates ``jaggy''
polyiamonds whose perimeter is not minimal.
Figures~\ref{fig:exmp_diamond}(a,b) illustrate this phenomenon.

\begin{figure}
  \centering
  \begin{subfigure}[b]{0.3\textwidth}
      \centering
      \drawpolyiamond[scale=0.4]{exmp_diamond.txt}
      \caption{$\poly$}
  \end{subfigure}
  \begin{subfigure}[b]{0.3\textwidth}
      \centering
      \drawpolyiamond[scale=0.4]{exmp_diamond_I.txt}
      \caption{$I(\poly)$}
  \end{subfigure}
  \begin{subfigure}[b]{0.3\textwidth}
      \centering
      \drawpolyiamond[scale=0.4]{exmp_diamond_II.txt}
      \caption{$\poly'$}
  \end{subfigure}
  \caption{An example of inflating polyiamonds.
            The polyiamond~$\poly$ is of a minimum perimeter, however, its inflated
            version, $I(\poly)$ is not of a minimum perimeter.
            The polyiamond~$\poly'$, obtained by adding to~$\poly$ all the cells
            sharing a \emph{vertex} with $\poly$, is a minimal-perimeter polyiamond.}
  \label{fig:exmp_diamond}
\end{figure}

However, we can fix this situation in the triangular lattice by modifying the definition
of the perimeter of a polyiamond so that it it would include all cells that share
a \emph{vertex} (instead of an edge) of the boundary of the polyiamond.
Under the new definition, Theorem~\ref{thm:main} holds.
The reason for this is surprisingly simple:
The modified definition merely mimics the inflation of animals on the graph dual to that
of the triangular lattice.  (Recall that graph duality maps vertices to faces (cells), and
vice versa, and edges to edges.)
However, the dual of the triangular lattice is the hexagonal lattice, for which we have
already shown in Section~\ref{sec:polyhexes} that all the premises of
Theorem~\ref{thm:main} hold.  Thus, applying the modified inflation operator in the
triangular lattice induces a bijection between sets of minimal-perimeter polyiamonds.
This relation is demonstrated in Figure~\ref{fig:exmp_diamond}.

\section{Conclusion}
\label{sec:conclusion}

In this paper, we show that the inflation operation induces a bijection between sets of
minimal-perimeter animals on any lattice which satisfies three conditions.
We demonstrate this result on three planar lattices:  the square, hexagonal, and also the
triangular (with a modified definition of the perimeter).
The most important contribution of this paper is the application of our result to polyhexes.
Specifically, the phenomenon of the number of isomers of a benzenoid hydrocarbons
remaining unchanged under circumscribing, which was observed in the literature of chemistry
more than~30 years ago but has never been proven till now.

However, we do not believe that this set of conditions is necessary.
Empirically, it seems that by inflating all the minimal-perimeter polycubes (animals on the
3-dimensional cubical lattice) of a given size, we obtain all the minimal-perimeter polycubes
of some larger size.  However, the second premise of Theorem~\ref{thm:main} does not hold for
this lattice.
Moreover, we believe that as stated, Theorem~\ref{thm:main} applies only to 2-dimensional
lattices! 
A simple conclusion from Lemma~\ref{lemma:pnc_size} is that if the premises of
Theorem~\ref{thm:main} hold for animals on a lattice~$\lattice$,
then~$\minp_\lattice(n) = \Theta(\sqrt{n})$.
We find it is reasonable to assume that for a $d$-dimensional lattice $\lattice_d$, the
relation between the size of a minimal-perimeter animal and its perimeter is roughly equal
to the relation between a $d$-dimensional sphere and its surface area.
Hence, we conjecture that $\minp^{\lattice_d}(n) = \Theta(n^{1-1/d})$, and, thus,
Theorem~\ref{thm:main} is not suitable for higher dimensions.



\bibliographystyle{splncs03}
\bibliography{references}

\end{document}